\documentclass[pra,twocolumn,floatfix]{revtex4}

\usepackage{amssymb,amsmath,amsthm,amsfonts}
\usepackage{graphicx}
\usepackage{subfigure}

\newcommand{\field}[1]{\mathbb{#1}}

\newcommand{\Z}{\field{Z}}

\newcommand{\ket}[1]{\ensuremath{| #1 \rangle}}

\newcommand{\braket}[2]{\ensuremath{\langle #1 | #2 \rangle}}
\newcommand{\set}[1]{\ensuremath{\left \{ #1 \right \}}}

\newcommand{\brackets}[1]{\ensuremath{\left( #1 \right)}}
\newcommand{\sbrackets}[1]{\ensuremath{\left[ #1 \right]}}

\begin{document} 

\title{Fermionized photons in the ground state of one-dimensional coupled cavities}

\author{Adam~G.~D'Souza, Barry~C.~Sanders and David~L.~Feder} 
\affiliation{Department of Physics and Astronomy and Institute for Quantum
Science and Technology, University of Calgary, Calgary, Alberta, Canada T2N 1N4}

\date{\today}

\begin{abstract}
The Density Matrix Renormalization Group algorithm is used to characterize the 
ground states of one-dimensional coupled cavities in the regime of low photon 
densities. Numerical results for photon and spin excitation 
densities, one- and two-body correlation functions, superfluid and condensate 
fractions, as well as the entanglement entropy and localizable entanglement are 
obtained for the Jaynes-Cummings-Hubbard (JCH) model, and are compared with 
those for the Bose-Hubbard (BH) model where applicable. The results indicate 
that a Tonks-Girardeau phase, in which the photons are strongly fermionized, 
appears between the Mott-insulating and superfluid phases as a function of the 
inter-cavity coupling. In fact, the superfluid density is found to be zero in 
a wide region outside the Mott-insulator phase boundary. The presence of two 
different species of excitation (spin and photon) in the JCH model gives rise 
to properties with no analog in the BH model, such as the (quasi)condensation 
of spin excitations and the spontaneous generation of entanglement between the 
atoms confined to each cavity.
\end{abstract}

\maketitle

\section{Introduction}
\label{sec:introduction}

The idea of simulating a complex, many-body physical system with a simpler 
model system has a long and rich history.  A prominent example is the 
Bose-Hubbard (BH) model~\cite{Fisher1989}, which governs the dynamics of 
bosons tunneling between sites of a lattice with energy $t$ and interacting 
on a given site with energy $U$. Originally proposed to describe the 
behavior of superfluid $^4$He in porous Vycor glass, it is now applied to a 
plethora of experimental systems including Josephson junction 
arrays~\cite{Zant-1992}, ultracold atoms in optical lattices~\cite{Bloch2005}, 
photonic crystals~\cite{Corrielli2013}, and arrays of coupled 
cavities~\cite{Hartmann2006,Greentree2006, Angelakis2007,Houck2012}. 

For repulsive interactions ($U>0$), 
there is a competition between delocalization due to the tunneling and the
tendency to localize due to the energy cost of multiple occupancy of a given
site. As a consequence, the model exhibits a quantum (zero-temperature) phase 
transition~\cite{Fisher1989}. In the strongly interacting (weak tunneling) 
regime~$t/U\ll 1$, the ground state is predicted to be a Mott 
insulator (MI)~\cite{Mott1937,Mott1949,Mott1982}, in which each site is 
occupied by an (identical) integer number of bosons; for $t/U\gg 1$, a 
superfluid (SF)
state results~\cite{Kapitza1938,Allen1938,Landau1941}, in which the bosons 
become completely delocalized. The transition between these phases was first
realized in Bose-Einstein condensates confined in three-dimensional optical 
lattice potentials~\cite{Greiner2002}, where the ratio $t/U$ was controlled by 
varying the well depth, and subsequently observed in many other cold-atom 
optical lattice experiments in one, two and three 
dimensions~\cite{Stoferle2004,Spielman2007,Gemelke2009,Bakr2010,Haller2010,Trotzky2010}.

There has been much recent interest in observing similar quantum phase 
transitions in cavity quantum 
electrodynamics~\cite{Hartmann2006,Greentree2006,Angelakis2007}. The principal 
motivation for employing these environments is the robustness of available 
technology for producing, manipulating and detecting photons. Unfortunately,
photons do not intrinsically interact. Various strategies have been proposed to
overcome this limitation, and a vast array of interesting many-body states 
have been conjectured to appear as a 
result~\cite{Hartmann2007,Rossini2007,Carusotto2009,Kiffner2010,Halu2013,Hayward2012,Schiro2012}. 
One model that has attracted particular attention is the 
Jaynes-Cummings-Hubbard (JCH) 
model~\cite{Hartmann2006,Hartmann2008a,Makin2008}, comprising a lattice of 
high-finesse optical cavities each containing one or more two-level atoms, with 
neighboring cavities coupled via the overlap of their evanescent modes. The 
(bosonic) photons can then be considered to `tunnel' from cavity to cavity. The 
atom interacts with the quantized electromagnetic field present within the 
cavity according to the Jaynes-Cummings model~\cite{Jaynes1963}, and photon 
interactions are generated via the photon-blockade 
mechanism~\cite{Birnbaum2005}. Importantly, the lattice polaritons that 
constitute the spin-photon elementary excitations of the JCH model are also
expected to undergo a phase transition from a Mott insulator to a
superfluid~\cite{Greentree2006,Hartmann2006,Angelakis2007,Rossini2007,Aichhorn2008,Koch2009,Schmidt2009,Pippan2009,Schmidt2010,Knap2010,Hohenadler2011}.
While early proposals for the experimental realization of coupled cavities 
involved nitrogen vacancies in diamond~\cite{Greentree2006}, self-assembled
quantum dots in photonic crystals~\cite{Na2008}, and trapped
ions~\cite{Ivanov2009,Mering2009}, more recent proposals favor circuit 
QED~\cite{Nunnenkamp2011,Wu2011,Houck2012,Schmidt2013}.

The close similarity between the properties of the BH and JCH models suggests 
that the polariton superfluid resulting from arranging cavities along a row 
should be rather peculiar. In one dimension, the interaction energy of free
bosons is strongly enhanced relative to their kinetic energy at low 
densities~\cite{Bloch2005}, a result that will be discussed in greater detail
in Sec.~\ref{subsec:Tonks}. The resulting Tonks-Girardeau
gas~\cite{Tonks1936, Girardeau1960} is described as the hard-core limit of the 
(integrable) Lieb-Liniger model for bosons with delta-function interactions in
one dimension~\cite{Lieb1963}. These hard-core bosons can be exactly mapped to 
non-interacting fermions~\cite{Girardeau1960}; for general densities the state 
is well-described within the framework of Luttinger liquid 
theory~\cite{Giamarchi2003}. The Tonks-Girardeau gas was first realized in 
ultracold atomic Bose gases confined in one-dimensional optical 
lattices~\cite{Paredes2004, Kinoshita2004}. More recently, the transition 
between a Luttinger liquid and Mott-insulating state was 
observed~\cite{Haller2010}. These results suggest that a one-dimensional 
arrangement of coupled cavities should be sufficient to induce the mobile 
photons to behave entirely as fermions. In fact, this possibility was noted 
previously for a dissipative model~\cite{Carusotto2009}. The central goal of 
the present investigation is to show that the photons are effectively 
fermionized in the ground-state of the JCH model. A secondary goal is to make
a careful, side-by-side comparison of the JCH and BH models in one dimension.

In this work, the ground states of the zero-temperature one-dimensional JCH and 
BH models in the vicinity of the MI phase boundary are obtained numerically 
using finite-system density matrix renormalization group (DMRG) methods. 
Several quantities are calculated that provide evidence for the nature of the 
states, including particle densities, one-particle and two-particle correlation 
functions, the superfluid fraction, the condensate fraction, and the 
entanglement entropy. Finite-size scalings are performed in order to infer the
value of these quantities in the thermodynamic limit. 

Two main conclusions can be drawn from our work. 
First, for small cavity couplings the ground state consists of a polariton
MI phase as expected, though with some interesting features not 
shared by single-component Bose systems. Second, the results clearly reveal
the strong fermionization of both the photons and spins in the JCH system 
throughout the so-called superfluid phase in the low-density limit. Detailed 
comparisons are made with the ground states of the BH model in the equivalent 
parameter regimes. The main conclusion that can be drawn is that superfluidity 
is weakly manifested in this phase, if it exists at all.

The manuscript is organized as follows. The JCH and BH models are reviewed in 
Sec.~\ref{sec:models}, and the DMRG methods used in the characterization of 
their ground states is described. This section also discusses the properties of 
the Tonks-Girardeau gas. The numerical results are presented in 
Sec.~\ref{sec:results}, and the discussion and conclusions are found in
Sec.~\ref{sec:conclusions}.

\section{Models and Methods}
\label{sec:models}

In this work we compare the properties of the one-dimensional 
Jaynes-Cummings-Hubbard (JCH) and Bose-Hubbard (BH) models, using density 
matrix renormalization group (DMRG) methods. This section briefly provides 
the background to these models and describes the numerical methods employed in
the calculations.

\subsection{JCH model}
\label{subsec:JCH}

The behavior of a single two-level atom, in the pseudospin representation, confined to a single high-finesse cavity is given 
by the Jaynes-Cummings Hamiltonian~\cite{Jaynes1963}, written within the 
rotating-wave approximation~\cite{Mandel1995} as
\begin{equation}
H_{\rm JC} = \omega_c \brackets{a^{\dag}a + \frac{1}{2}} 
+ \frac{1}{2}\omega_a \sigma^z + g \brackets{a^{\dag} \sigma^- 
+ a \sigma^+}.
\label{eq:JC}
\end{equation}
Here, $\omega_c$ is the natural cavity frequency ($\hbar=1$ in this work for
convenience), $\omega_a$ is the excitation frequency of the atom, $a$ 
($a^{\dag}$) is the photon annihilation (creation) operator, $\sigma^z$ is 
the spin-$1/2$ representation of the Pauli $z$ operator, $\sigma^{\pm}$ are 
the spin raising and lowering operators, and $g$ is the strength of the 
atom-photon coupling, proportional to the magnitude of the inner product 
between the dipole vector and the local field. In this work, $g$ is assumed to
be a real quantity, equivalent to assuming that the dipole and field oscillate
in phase. This model describes an isolated system which ignores environmental
couplings. The cavity is therefore assumed to have arbitrarily high finesse, 
and be in the strong-coupling limit.

In the rotating-wave approximation the total number of excitations
\begin{equation}
N_i = a_{i}^{\dag}a_i+\sigma^{+}_i \sigma^{-}_i,
\label{eq:JCnumber}
\end{equation} 
is a conserved quantity (note that the spin number operator would normally 
contribute the term $\sigma^{-}_i \sigma^{+}_i$, but this corresponds to the population of atoms in the ground state; hence, it is not included in the \textit{excitation} number operator). The eigenstates of $H_{\mathrm{JC}}$ are coherent 
superpositions of photonic and spin excitations with a definite total 
excitation number, known as polaritons~\cite{Hartmann2006}. Within a particular 
excitation number block $N$, the eigenstates are given (c.f.\ Appendix A of 
Ref.~\cite{Koch2009}) by
\begin{align}
\label{eq:polaritonminus} \ket{N-} = \sin{\theta_N} \ket{Ng} + \cos{\theta_N} \ket{(N-1) e}; \\
\label{eq:polaritonplus} \ket{N+} = \cos{\theta_N} \ket{Ng} - \sin{\theta_N} \ket{(N-1) e},
\end{align}
with mixing angle
\begin{equation}
\theta_N = \frac{1}{2} \arctan{\brackets{\frac{2g\sqrt{N}}{\delta}}}.
\label{eq:polaritonmixing} 
\end{equation}
Here $\delta := \omega_c - \omega_a$ is the detuning of the cavity and atomic
frequencies. The eigenenergies of Eq.~(\ref{eq:JC}) for $N \geq 1$ are given by 
\begin{equation}
E_{N\pm} = N\omega_c + \frac{\delta}{2} \pm \left[ \left( \frac{\delta}{2}
\right)^2 + Ng^2 \right]^{1/2},
\label{eq:eigsJC}
\end{equation}
while for $N=0$, $E_0=0$. The energy levels are thus arranged in 
two-dimensional manifolds labeled by the polariton number $N$ (except for the 
$N=0$ sector, which is one-dimensional), separated by the energy of the 
single-photon cavity mode, $\omega_c$.

The anharmonicity in the eigenenergies~(\ref{eq:eigsJC}) of size $\sqrt{N}$ 
is the origin of the photon-blockade effect~\cite{Birnbaum2005}, giving rise
to effective photon interactions. Consider for simplicity the zero-detuning 
case $\delta=0$ giving eigenenergies $E_{N\pm}=N\omega_c\pm g\sqrt{N}$. With 
one photon the cavity has the lower energy eigenvalue 
$E_{1-}=\omega_c-g$. Na\"\i vely, two independent photons would yield the total 
energy $2E_{1-}=2\omega_c-2g$, but in fact the two-photon energy eigenvalue is 
$E_{2-}=2\omega_c-g\sqrt{2}$. The difference between these energies yields an
estimate for the effective repulsive photon-photon interaction strength:
$E_{2-}-2E_{1-}=(2-\sqrt{2})g$.

In the Jaynes-Cummings-Hubbard (JCH) model, the cavity mode leakage is no 
longer neglected. Instead, one imagines a regular lattice of $L^d$ cavities in $d$ dimensions positioned sufficiently 
close together that a photon emitted from one cavity can be absorbed into an 
adjacent cavity with energy (rate) $\kappa$. The JCH model is written as
\begin{equation}
H_{\mathrm{JCH}} = \displaystyle \sum_i \brackets{H_{\mathrm{JC},i} -\mu N_i} 
- \kappa \sum_{\langle i,j \rangle} \brackets{a^{\dag}_i a_j + a^{\dag}_j a_i},
\label{eq:JCH1}
\end{equation}
where $H_{{\rm JC},i}$ and $N_i$ are Eqs.~(\ref{eq:JC}) and (\ref{eq:JCnumber})
respectively with $\{a,a^{\dag},\sigma^z,\sigma^{\pm}\}$ replaced by
$\{a_i,a^{\dag}_i,\sigma^z_i,\sigma^{\pm}_i\}$, and $i$ labels the position of 
a cavity. The $\langle i,j\rangle$ notation indicates that the sum is over
nearest neighbors. The chemical potential $\mu$ fixes the mean polariton 
number, and is employed primarily to connect with the results of the BH model 
which is generally solved in the grand canonical ensemble. Note that in the 
JCH model, the atoms are fixed within the cavities, and only the photons are 
able to `tunnel' from cavity to adjacent cavity.

For reasons that will become clearer momentarily, it is convenient to rescale
the JCH energy in units of the coupling constant $g$; in the limit of zero
detuning $\omega_c=\omega_a$ one can rewrite the Hamiltonian~(\ref{eq:JCH1}) as
\begin{eqnarray}
\frac{H_{\rm JCH}}{g}&=&-\frac{\kappa}{g}\sum_{\langle i,j\rangle}\left(
a_i^{\dag}a_j+a_j^{\dag}a_i\right)\nonumber \\
&+&\sum_i\left[a_i^{\dag}\sigma_i^-+a_i\sigma_i^+
-\left(\frac{\mu-\omega_c}{g}\right)N_i\right],
\end{eqnarray}
where unimportant additive constant terms are omitted.
The first term corresponds to the photon hopping, the second to the local JC 
term, and the last term can be considered as a rescaled chemical potential for 
the total polariton density $\tilde{\mu}:=(\mu-\omega_c)/g$.

\subsection{BH Model}
\label{subsec:BH}

In the BH model, bosons tunnel between nearest neighboring sites of a lattice, 
and experience on-site interactions (which can be either attractive or 
repulsive in general). The model is described by the Hamiltonian
\begin{equation}
H_{\mathrm{BH}} =-t\sum_{\langle i,j \rangle} \brackets{b^{\dag}_i b_j 
+ b^{\dag}_j b_i} 
+\sum_i\left[\frac{U}{2}N_i (N_i - 1) - \mu N_i\right],\
\label{eq:BH}
\end{equation}
where $b_i$, $b^{\dag}_i$, and $N_i:=b^{\dag}_ib_i$ are respectively the 
bosonic annihilation, creation, and number operators for site $i$, $t>0$ 
is the nearest-neighbor tunneling amplitude and $U$ is the on-site interaction 
energy. The chemical potential fixes the mean boson density on the lattice. In
this work only repulsive interactions $U>0$ will be considered.

While a direct mapping between the JCH and BH models is not possible because 
the former has two different kinds of excitations while the latter has only
one, the parameters can be chosen in such a way as to simplify comparisons.
One can rescale the BH energies in terms of the interaction strength by 
dividing Eq.~(\ref{eq:BH}) by $U$. In this case the hopping amplitudes are
$\kappa/g$ and $t/U$ in the JCH and BH models, respectively. The effective
interaction strength between photons is $(2-\sqrt{2})g$, which implies that the 
two systems should become similar for $t/U\sim\kappa/(2-\sqrt{2})g
=\left(1+\frac{1}{\sqrt{2}}\right)(\kappa/g)\approx 1.707(\kappa/g)$. A similar
connection can be obtained between the chemical potentials of the two models:
$\tilde{\mu}+1\sim(2-\sqrt{2})\mu/U$ or $\mu/U\sim 1.707(\tilde{\mu}+1)$.
Note that these scalings are valid only when only a single atom is confined to 
each cavity.

As discussed in the Introduction, in the weak tunneling (strong interactions)
limit $t/U\ll 1$ the ground state is an incompressible MI
characterized by localized bosons, (constant) integer occupation of a given 
site, and an energy gap to excitations of order $U$. Deep in this limit, the 
ground-state wavefunction can be approximated as 
$|\Psi\rangle\sim\prod_ib_i^{\dag}
|\Phi\rangle$, where $|\Phi\rangle$ is the particle vacuum state and the 
product is over all lattice sites. Because each site is independent of any 
other, the overlap of the states $b_s|\Psi\rangle$ and $b_r|\Psi\rangle$ is 
exactly zero unless $r=s$. The one-body boson correlation function
\begin{equation}
G^{(1)}(r,s)=\langle b_r^{\dag}b_s\rangle
\label{eq:G1}
\end{equation}
and the two-body correlation function
\begin{equation}
G^{(2)}(r,s)=\langle b_r^{\dag}b_s^{\dag}b_sb_r\rangle
\label{eq:G2}
\end{equation}
will then be zero for all $r\neq s$. In reality, for any finite $t/U$ the 
gapped ground state will deviate from this simple prediction and the 
correlation functions should instead decrease exponentially in $|r-s|$ with a 
characteristic length scale $\xi\sim 1/U$ that scales as the inverse of the 
gap to excitations~\cite{Hastings2006}. The correlation length diverges as a 
system becomes critical~\cite{Sachdev2000}. One would therefore expect $\xi$ to 
increase from $0$ to $\infty$ as the hopping goes from 0 to its critical value 
at the phase boundary for a fixed chemical potential. 

The phase boundary in $\mu$-$t$ space, known as the `Mott lobe,' is roughly 
semi-circular in profile in two and three dimensions~\cite{Fisher1989}. For
$\mu\notin\Z$, the system remains in the MI phase with increasing $t$ until
some critical value at which point the system undergoes a phase transition to
SF; likewise for constant $t$ and increasing $\mu$. The Mott lobe becomes 
strongly distorted in one dimension~\cite{Kuehner1998}, and the system displays 
re-entrance: at constant $\mu$, on increasing $t$ the ground state phase 
changes from MI to SF to MI and back to SF again.

In the strong tunneling (weak interactions) limit $t/U\gg 1$, the ground state 
of the BH model corresponds to an interacting Bose-Einstein condensate. Each 
boson is highly extended throughout the lattice, and the ground state can be 
approximated by $|\Psi\rangle\sim\left(\sum_ib_i^{\dag}\right)^{N_B}|\Phi
\rangle$, where $N_B$ is the number of bosons. This 
compressible state is characterized by a gapless linear spectrum and long-range 
correlation functions~(\ref{eq:G1}) and (\ref{eq:G2}) that are independent of 
$|r-s|$. In one-dimension, however, true Bose-Einstein condensation is not
possible; rather, the ground state corresponds to a quasi-condensate 
with only algebraic long-finite order and characterized by strong 
fluctuations~\cite{Popov1987}. Instead one finds 
$G^{(1)}(r,s)\sim 1/|r-s|^{\alpha}$, where the parameter $\alpha$ characterizes
the degree of quasi-condensation ($\alpha\to 0$ for a true condensate).

\subsection{Tonks-Girardeau Gas}
\label{subsec:Tonks}

At very low densities, one-dimensional repulsively interacting bosons form a 
Tonks-Girardeau gas, and effectively behave as non-interacting 
fermions~\cite{Girardeau1960}. In the absence of any external potential (other
than the ones used for confinement), the ground state properties are governed
by the kinetic and interaction potential energies $T$ and $U$. The mean kinetic 
energy per particle scales as $T/N\sim1/m\ell^2
\sim \langle n\rangle^2/m$, where $\ell$ is the mean interparticle 
distance which in one dimension scales as the inverse of the mean particle 
density $\ell\sim 1/\langle n\rangle$. (In the presence of a weak lattice, the 
bare boson mass $m$ is rescaled to an effective mass $m^*\sim 1/t$). 
When the interaction potential can be modeled in terms of a pseudopotential 
(low energy, long-wavelength collisions), one can write the mean interaction 
potential in one dimension as $U/N\sim\langle n\rangle
/m|a_{\rm 1D}|$~\cite{Olshanii1998}, where $a_{\rm 1D}$ is the one-dimensional 
s-wave scattering length. The Tonks parameter, the ratio of the potential and 
kinetic energies $\gamma=U/T=2/\langle n\rangle|a_{\rm 1D}|$ is therefore huge 
at low densities, in marked contrast to the situation in higher dimensions. To 
minimize the interaction potential, particles prefer to be as far apart from 
one another as possible, much like fermions.

The free fermionic wavefunction can be written in terms of a Slater 
determinant to guarantee the proper antisymmetrization of the wavefunction. For
example, a system of $N$ free fermions on $L$ sites has a wavefunction given in 
the position representation by
\begin{equation}
\Psi_F(r_1, \dots, r_L) = \text{det} \left[
\begin{array}{cccc}
\phi_1(r_1) & \phi_1(r_2) & \dots & \phi_1(r_L) \\
\phi_2(r_1) & \phi_2(r_2) & \dots & \phi_2(r_L) \\
\vdots & \vdots & \ddots & \vdots \\
\phi_N(r_1) & \phi_N(r_2) & \dots & \phi_N(r_L)
\end{array}
\right],
\end{equation}
where the $\set{r_i}$ indicate the positions of the lattice sites and the 
$\set{\phi_i}$ are single-particle wavefunctions. In the perfectly hard-core 
limit of the Tonks-Girardeau gas, the ground state of the fermionized bosons 
is simply
\begin{equation}
\Psi_B(r_1, \dots, r_L)= \prod_{i<j}^L{\rm sgn}(r_i-r_j)\Psi_F(r_1, \dots, r_L),
\label{eq:TonksPsi}
\end{equation}
where the factor multiplying $\Psi_F$ ensures that all negative signs
associated with the interchange of two fermions disappears. 

Many properties are
shared by $\Psi_F$ and $\Psi_B$. For example, the local density profile of both 
systems in real space is the same, since $|\Psi_B(r_1, \dots, r_N)|^2
= |\Psi_F(r_1, \dots, r_N)|^2$~\cite{Yukalov2005}. Similarly, all density
correlation functions are the same~\cite{Cazalilla2011}; for example, for a 
ring of length $L\to\infty$, the normalized two-body correlation function is
\begin{equation}
g^{(2)}(r,s)=\frac{\langle b_r^{\dag}b_s^{\dag}b_sb_r\rangle}
{\langle b_r^{\dag}b_r\rangle\langle b_s^{\dag}b_s\rangle}
= 1-\left[\frac{\sin(\pi n|r-s|)}{\pi n|r-s|}\right]^2,
\label{eq:g2Tonks}
\end{equation}
where $n$ is the mean particle density. The correlation function is zero at 
$r=s$, reflecting the Pauli exclusion principle; this behavior is referred to 
as the `exclusion hole'. Away from this point the correlation function grows 
and displays Friedel oscillations~\cite{Friedel1958} that decay with increasing 
$|r-s|$. For one-dimensional spinless fermions, the oscillations have 
wavelength $\lambda_F=1/n=2\pi/k_F$ where $k_F$ is the Fermi wavevector. Thus,
the presence of an exclusion hole and Friedel oscillations in the two-body
correlation function is a `smoking gun' for the fermionization of bosons in
the Tonks-Girardeau gas.

For a finite system with $N$ free fermions on $L$ sites with open boundary
conditions, such as is considered in this work, a straightforward calculation 
yields
\begin{equation}
\label{eq:g2FF} G^{(2)}(r,s) = \brackets{\frac{N}{L+1}}^2\sbrackets{B(r,s) 
- A(r,s)},
\end{equation}
where
\begin{align}
A(r,s)&=\left[\frac{\cos{\frac{\pi\brackets{N+1}\brackets{r-s}}{2(L+1)}}
\sin{\frac{\pi N\brackets{r-s}}{2(L+1)}}}
{N\sin{\frac{\pi\brackets{r-s}}{2(L+1)}}}\right.
\nonumber \\
&-\left.\frac{\cos{\frac{\pi\brackets{N+1}\brackets{r+s}}{2(L+1)}}
\sin{\frac{\pi N\brackets{r+s}}{2(L+1)}}}
{N\sin{\frac{\pi\brackets{r+s}}{2(L+1)}}}\right]^2;\nonumber \\
B(r,s)&=\sbrackets{1-\frac{\cos{\frac{\pi\brackets{N+1}r}{(L+1)}}}{N}
\frac{\sin{\frac{\pi Nr}{(L+1)}}}{\sin{\frac{\pi r}{(L+1)}}}}\nonumber \\
&\times\sbrackets{1-\frac{\cos{\frac{\pi\brackets{N+1}s}{(L+1)}}}{N}
\frac{\sin{\frac{\pi Ns}{(L+1)}}}{\sin{\frac{\pi s}{(L+1)}}}}\label{eq:g2FF2}.
\end{align}
It is simple to verify that $G^{(2)}(r,r)\to 0$. For large separations between 
particles $\left|\frac{(r-s)}{(L+1)}\right| \gg 0$, one finds that $A(r,s)$ 
and $B(r,s)$ oscillate in the vicinity of zero and unity, respectively, so that 
$G^{(2)}(r,s) \approx \brackets{\frac{N}{L+1}}^2.$ Choosing the location of 
one particle at the center of the chain $r=\lceil L/2 \rceil$, $G^{(2)}$ far 
from the center oscillates about a mean value approximately equal to the 
square of the mean particle density $n$. For $\left|\frac{r-s}{L+1}\right|
\gg 0$, the oscillation of $G^{(2)}$ is governed by the last term in the 
definition of $B(r,s)$ in Eq.~(\ref{eq:g2FF2}). In the thermodynamic limit 
$N,L\to\infty$ but $n=N/L\to$~const., one obtains $B(\lceil L/2 \rceil,s)
\approx 1-\sin(2\pi ns)/2\pi ns$. The Friedel oscillation wavelength is 
therefore again $\lambda_F=1/n=2\pi/k_F$.

The single-body correlation function is not the same for the Tonks-Girardeau 
and free fermion gases, however: the sign function in Eq.~(\ref{eq:TonksPsi}) 
does not
disappear when inserted into Eq.~(\ref{eq:G1}). The calculation of this 
quantity is quite involved~\cite{Cazalilla2011}, but the asymptotic behavior 
$|r-s|\gg 0$ but $|r-s|\ll L$ is found to be
\begin{equation}
G^{(1)}(r,s)\sim\frac{1}{\sqrt{2n_0L|\sin(\pi|r-s|/L)}}.
\label{eq:G1Tonks}
\end{equation}
For $|r-s|\ll L$ one obtains $G^{(1)}(r,s)\sim 1/\sqrt{|r-s|}$, which indicates
that for the Tonks-Girardeau gas the exponent of the power law is $\alpha=1/2$.
Another `smoking gun' for the Tonks-Girardeau phase is therefore the 
power-law behavior of the one-body density matrix with exponent $\alpha=1/2$.

The Fourier transform of the one-particle correlation function $G^{(1)}(r)$ is 
the momentum distribution $n(k)$. For the Tonks-Girardeau gas, the power-law 
behavior at long distances translates into a power-law divergence of the 
momentum 
distribution at long wavelengths, $n(k)\sim 1/|k|^{1/2}$ for $k\to 0$. This 
highly peaked distribution is reminiscent of the delta-function distribution 
that one would expect if the bosons formed a Bose-Einstein condensate, except
it is now broadened due to the finite-range phase order associated with the 
quasi-condensation. This distribution is dramatically different from that of
a non-interacting Fermi gas, where $n(k)$ is a constant for all $k \leq k_F$
and is zero otherwise ($k_F$ is the Fermi wavevector). The momentum 
distribution for a Tonks-Girardeau gas in a weak axial trapping potential has
been experimentally observed~\cite{Paredes2004,Kinoshita2004}.

\subsection{Numerical Methods}
\label{subsec:methods}

The characteristics of the BH and JCH models were obtained by means of 
finite-system density matrix renormalization group (DMRG) simulations. We 
employed the DMRG code from the Algorithms and Libraries for Physics 
Simulations (ALPS) project~\cite{Albuquerque2007,Bauer2011}. Simulations were 
carried out for systems of size $L=15,19,23,27,31$ with both open boundary
conditions (equivalent to hard-wall boundary conditions) and periodic boundary 
conditions, and a finite-size scaling analysis was performed for all quantities 
(unless explicitly noted) in order to infer the results for the thermodynamic 
limit. We use DMRG as the method because it is suitable for obtaining results 
that are so precise as to be considered exact~\cite{White1993}, while being 
able to handle much larger finite-size systems than exact 
diagonalization~\cite{Schollwock2005,Schollwock2011}.

The bulk of the simulations employed open boundary conditions, in order to
accelerate convergence. For the BH (JCH) model, a maximum of $N_{\text{max}}=5$ 
$(6)$ bosons (photons) per site (cavity) were allowed, and we kept $M=80$ 
(100) states. For the JCH model, this corresponds to a Hilbert dimension 
$D=12$ for each cavity. For the superfluid fraction, the method chosen 
necessitated the use of 
periodic boundary conditions. Usually the number of states kept for these
simulations is on the order of the square of the number chosen for open 
boundary conditions; however, since the method only required ground state 
energies and not correlation functions, the numerical requirements were not as 
stringent. The superfluid fraction calculation for the BH (JCH) model was 
performed using $N_{\text{max}}=7$ $(6)$ and $M=200$ (140). In all cases, we 
verified that increasing the values of $M$ and $N_{\text{max}}$ did not 
change the ground state energies or correlation functions. For the BH (JCH)
system, these parameters correspond to a maximal Hilbert space dimension of 
$1.60\times10^5$ $(1.44\times10^6)$ for the simulations with open boundary 
conditions, and $1.96\times 10^6$ $(2.82 \times 10^6)$ for periodic boundary 
conditions. 

We used eight finite-size sweeps for all simulations, and verified that the 
ground state energy and correlation functions did not change by increasing the 
number of sweeps. The calculation times for a single run for the simulations 
for the BH (JCH) models were typically approximately 1 hour (24 hours) when
using open boundary conditions, while the runs using periodic boundary 
conditions required up to approximately 8 (24) hours. In order to keep the 
calculation time to a minimum, we assumed that the total boson (polariton) 
number was a conserved quantity in every case. Note that this does not pose a 
problem even in the superfluid phase because the superfluid density need not 
correspond to the mean boson (polariton) density.

Parameters for the hopping ($t$ or $\kappa$ for the BH or JCH models,
respectively) and chemical potential $\mu$ were chosen in order to remain in 
the vicinity of the MI phase boundary. In 1D, the tip of the Mott 
$n=1$ lobes in the $(\mu/U,t/U)$ and $(\tilde{\mu},\kappa/g)$ planes for 
the BH and JCH models are found to be located at approximately 
$(0.09,0.3)$~\cite{Kuehner1998,Kuehner2000,Ejima2011} and approximately
$(-0.95,0.2)$~\cite{Rossini2007,Mering2009}, respectively. This value of 
$\kappa/g$ is consistent with the rescaling factor of approximately $1.7$ 
between the BH and JCH models, discussed in Sec.~\ref{subsec:BH}. To span most 
of the Mott lobe, the range of hopping is therefore chosen to be 
$t/U\in[0,0.27]$ and $\kappa/g\in[0,0.16]$ for the two models. Likewise, the 
phase boundaries for $t=0$ and $\kappa=0$ correspond to 
$\mu/U=n$~\cite{Fisher1989} and 
$\tilde{\mu}=\sqrt{n}-\sqrt{n+1}$~\cite{Koch2009} for the BH and JCH 
models in any dimension, respectively. Thus, the transition from the $n=0$ to 
$n=1$ Mott lobes at zero hopping occurs for $\mu/U=1$ and $\tilde{\mu}=-1$
for the two models; the transition to the $n=2$ Mott lobe occurs for $\mu/U=2$
and $\tilde{\mu}\approx -0.41$. To capture some of the $n=0$ lobe and
approximately half of the $n=1$ lobe, we chose chemical potentials in the range
$\mu/U\in[-0.17,0.55]$ and $\tilde{\mu}\in[-1.10,-0.68]$. Only the
simplest zero-detuning case $\delta=0$ is considered in this work. Previous
work has shown that detuning can be a useful parameter, changing the effective
strength of interactions and thereby the phase 
diagrams~\cite{Koch2009,Schmidt2009,Schmidt2010,Schmidt2013}.

\section{Results}
\label{sec:results}

\subsection{Density phase diagrams}
\label{subsec:densities}

\begin{figure}[t]
	\subfigure[][]{	
		\includegraphics[height=0.172\textwidth]{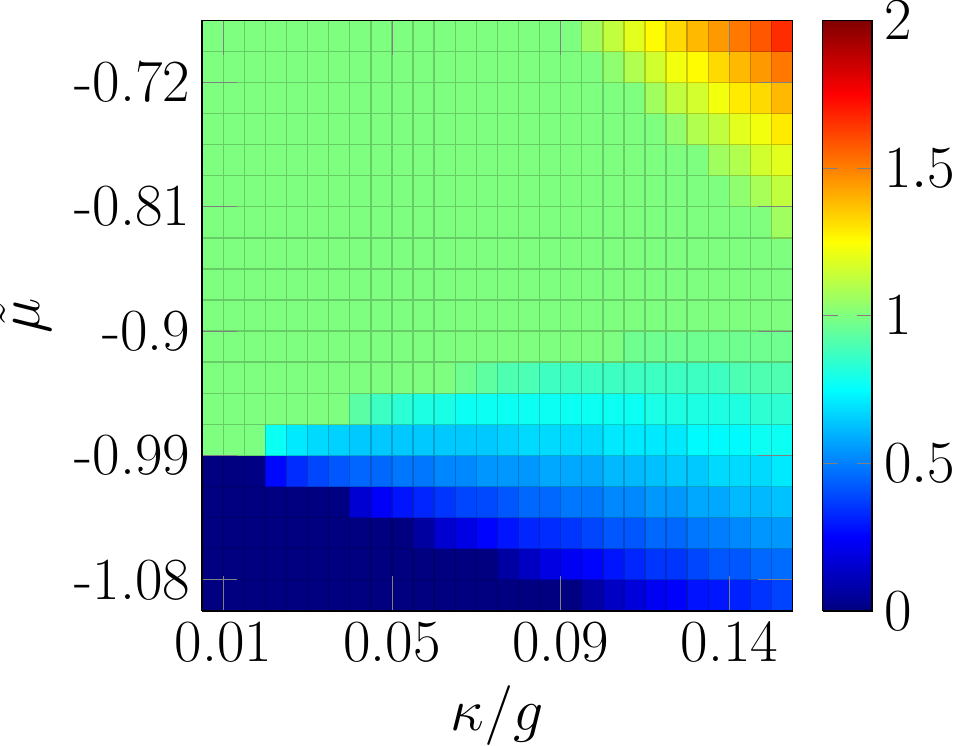}
		\label{JCH-density-polaritons-31}		
	}
	\subfigure[][]{
		\includegraphics[height=0.172\textwidth]{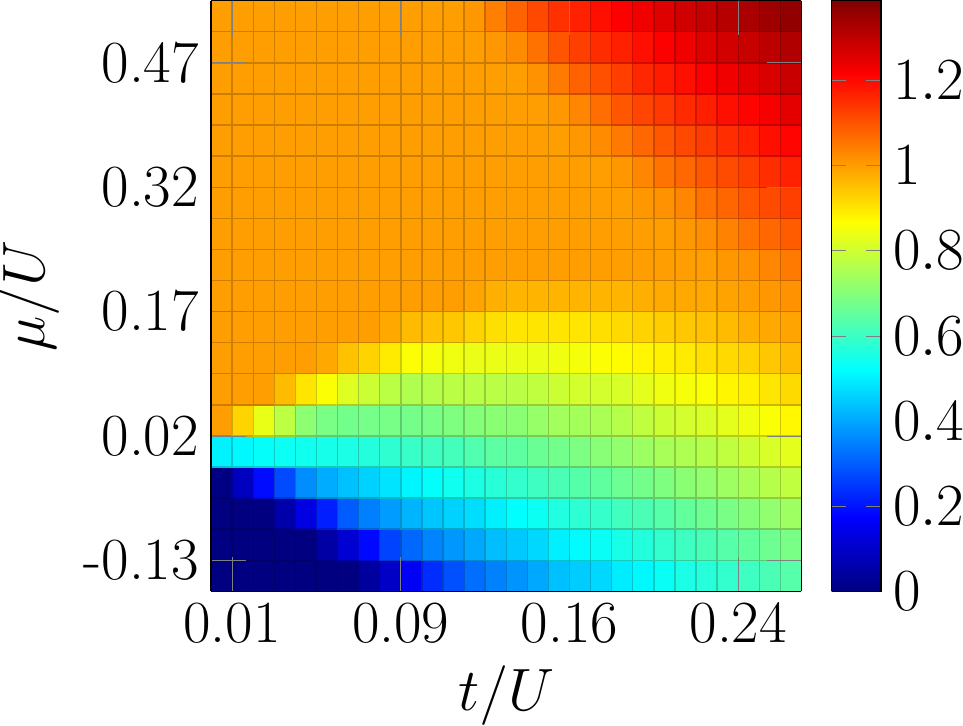}
		\label{BH-density-31}
	}
\caption{Mean density phase diagrams for (a)~polaritons in the 1D JCH model and 
(b)~bosons in the 1D BH model, in both cases for a system size $L=31$. Regions
of constant mean density correspond to MI states.}
	\label{fig:densities}	
\end{figure}

The location of the phase boundary between the gapped MI phase and the SF 
phase has been previously established numerically in the 
thermodynamic limit with finite-size DMRG, for both the 1D Bose-Hubbard 
model~\cite{Kuehner1998,Kuehner2000,Ejima2011} and the 1D JCH model~\cite{Rossini2007}. 
The results for the mean densities of bosons and polaritons are shown in 
Fig.~\ref{fig:densities} for the parameter sets discussed in the previous 
section and the largest number of lattice sites $L=31$ studied. Open boundary
conditions are employed in this case; for periodic boundary conditions the
density on any site would coincide with the mean density. While such
density plots have not to our knowledge been previously shown in the 
literature, the main 
purpose of showing these plots here is to orient the reader to the location in 
phase space for which the simulations have been conducted. The ranges of the 
normalized hopping parameters $t/U, \kappa/g$ and effective chemical potentials
$\mu/U, \tilde{\mu}$ are chosen to be equivalent, based on the scaling 
assumption given in Sec.~\ref{subsec:BH}. For both models, the goal is to 
explore the regions in the vicinity of the transition between the MI and SF 
phases. Unlike most previous studies, this work focuses particularly on the 
low-density region where fermionization is expected.

The mean densities of either bosons or polaritons can be used to distinguish 
the two phases in both models, since the mean density is pinned to an integer 
in the MI regime, but not in the SF regime. Consequently, the locations of the 
phase boundary for our finite-size systems are quite clearly visible in 
Fig.~\ref{fig:densities}. The shapes of the phase boundaries closely resemble
the thermodynamic limit results of Refs.~\cite{Kuehner1998,Rossini2007},
though their positions are shifted slightly due to the finite-size system. 
Depicted is the region of the phase diagram where the $n=0$ and $n=1$ Mott 
lobes meet, as well as the low-density superfluid regions near the boundaries 
of these lobes. Roughly, the mean density of bosons (polaritons) increases with 
increasing $t$ ($\kappa$). This work is mainly concerned with the low-density
superfluid regions of phase space, corresponding to low hopping and chemical
potential.

The re-entrant shape of the $n=1$ Mott lobe can be clearly seen as the 
constant-density region of the BH model in Fig.~\ref{BH-density-31}, but is not 
as obvious in 
Fig.~\ref{JCH-density-polaritons-31}. At certain fixed values of $\mu$ within a 
continuous range, monotonically increasing the hopping parameter $t$ or 
$\kappa$ from zero causes the system to transition from the MI to the SF, back 
to the MI and again back to the SF regime~\cite{Rossini2007}. Viewed from 
within the Mott lobe, the lower part of the phase boundary (the hole boundary) 
is concave, while the upper part (the particle boundary) is convex. The 
particle and hole boundaries meet at a sharply cusped tip. The phase transition along the line of constant mean density passing through this point is in the $(d+1)$-dimensional XY universality class, which for $d=1$ is of the Berezinskii-Kosterlitz-Thouless (BKT) 
type~\cite{Fisher1989,Kosterlitz1973} with Tomonaga-Luttinger parameter 
$K_b=1/2$~\cite{Kuehner2000,Ejima2011}. The MI-to-SF transition across either 
the particle or the hole boundary is generic (i.e.\ Gaussian like the 
condensation transition of an ideal Bose gas) and characterized by 
$K_b=1$~\cite{Kuehner2000}. This implies that the SF phase near the 
particle or hole boundaries should be characterized by one-particle 
correlation functions $G^{(1)}(r)\sim|r|^{-K_b/2}\sim 1/r^{1/2}$, consistent 
with the Tonks-Girardeau gas scaling.

\begin{figure}[t]
	\subfigure[][]{
		\includegraphics[height=0.175\textwidth]{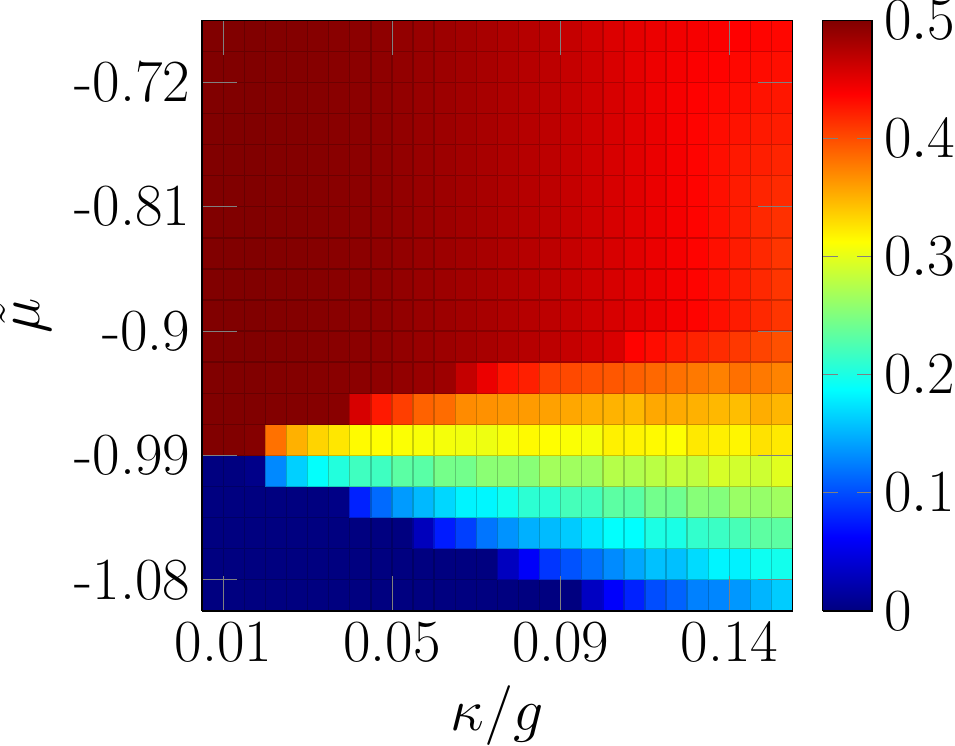}
		\label{fig:atomdensities-31}
	}
	\subfigure[][]{
		\includegraphics[height=0.175\textwidth]{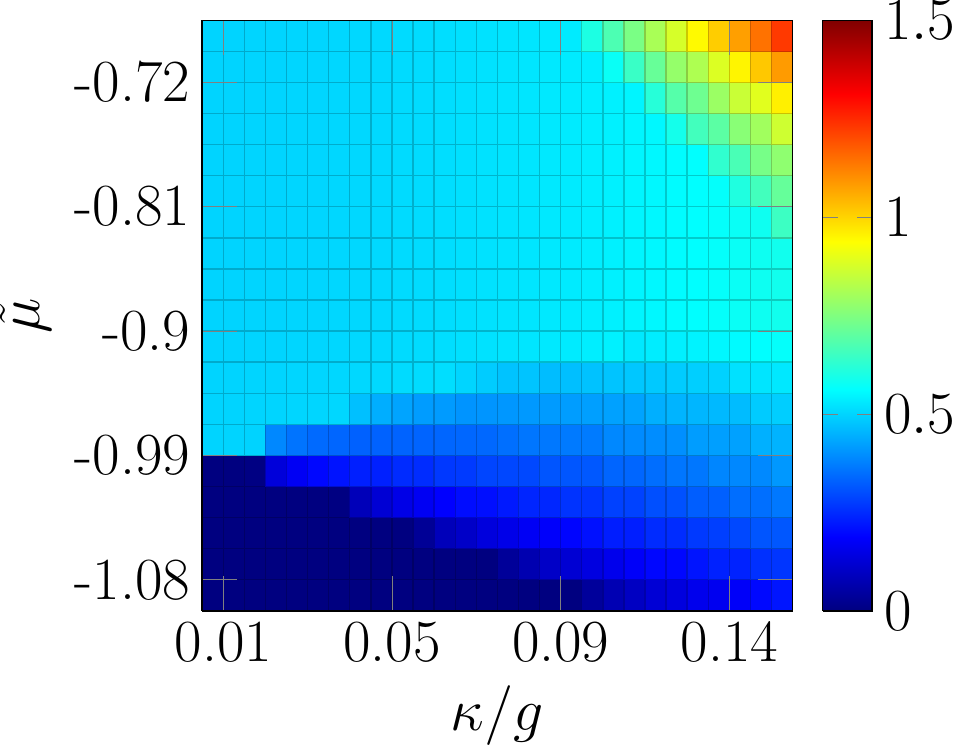}
	}
\caption{(a) Mean spin and (b) photonic excitation densities in the JCH model 
with $L=31$. These can vary throughout the MI phase.}
	\label{fig:atomphotondensities-31}
\end{figure}

In the JCH model, the mean densities of the spin and photonic species, 
depicted in Fig.~\ref{fig:atomphotondensities-31}, need not remain proportional
to track each other. Consider first the mean spin excitation density, 
Fig.~\ref{fig:atomdensities-31}. In the 
atomic limit $\kappa=0$, the cavities are decoupled from each other and thus 
the overall ground state is the $L$-fold tensor product of the single-cavity 
lower-lying polariton states
\begin{equation}
\displaystyle \ket{\psi_{\text{ground}}}=\bigotimes_{i=1}^L \ket{1-}_i. 
\end{equation}
From Eqs.~(\ref{eq:polaritonminus}) and (\ref{eq:polaritonmixing}) in the case 
of zero-detuning, the single-cavity polariton ground states $\ket{1-}$ are 
equal-weight superpositions of a photonic and a spin excitation. Consequently, 
the mean spin excitation density is equal to $\frac{1}{2}$ in this limit. 
Conversely, in the hopping-dominated limit 
$\kappa \rightarrow \infty$, the photons and spins decouple. Each atom is then
in the ground state so the mean spin excitation density vanishes. At 
intermediate $\kappa$ between these extremes, the data in 
Fig.~\ref{fig:atomphotondensities-31} show that across the lower boundary of 
the Mott lobe (the hole boundary) there is a sharp drop in the mean density of 
spin excitations, which can be used to distinguish the two phases. By 
contrast, crossing the upper boundary, the mean spin excitation density 
varies smoothly from $\frac{1}{2}$ in the atomic limit to lower values, 
presumably tending towards $0$ in the limit $\kappa \rightarrow \infty$.

Next consider the mean photon density. Unlike the spin excitations, the 
number of photons and hence the mean photon density is unbounded from above in 
the grand canonical ensemble. In the large-hopping limit at fixed chemical 
potential, one expects the mean photon density to increase. In fact, for a 
lattice with coordination number $z_c$, an instability occurs when the hopping 
roughly satisfies $z_c\kappa/g > -\tilde{\mu}$; for larger values of the 
hopping, the ground state energy decreases without bound as a function of 
increasing photon number~\cite{Koch2009}. This regime is not considered in the 
calculations, as it corresponds to larger densities deep in the SF phase where
fermionization is unlikely. The mean photon density within the $n=1$ MI lobe 
tracks with the mean spin excitation density in such a way that the overall 
polariton 
density is pinned to 1 per site, as expected. Crossing the upper (particle) 
boundary of the Mott lobe, the mean density of photons begins to increase 
rapidly with increasing hopping. This indicates that in the hopping-dominated 
limit, the system behaves like a photon superfluid, with the effects of the 
spins becoming negligible. On the other hand, in the intermediate regions 
between the $n=0$ and $n=1$ lobes, the mean photon density remains low; the 
reason is that at low hopping in one dimension, the effective repulsive 
interactions between photons become strong and thus there is an energy cost 
associated with adding photons to the system.

\subsection{Correlation functions}
\label{subsec:correlations}

\subsubsection{One-body density matrix}
Consider now the single-particle correlation function $G^{(1)}$, defined in 
Eq.~(\ref{eq:G1}). This has been calculated previously via DMRG for the 1D BH 
model, to verify the asymptotic predictions of the Luttinger liquid 
theory~\cite{Pai1996}, and to estimate the location of the critical value of 
$t/U$ for the BKT transition~\cite{Kuehner1998}. Similar plots of $G^{(1)}$ for 
varying interaction strengths in the $n=1$ MI lobe of the 1D BH model are 
shown in Ref.~\onlinecite{Ejima2012}. The 
normalized version $C_r(s-r):=G^{(1)}(r,s)/\sqrt{N_rN_s}$ was also considered
for the 1D BH model with an additional harmonic trapping 
potential~\cite{Kollath2004} for various different sites $r$, and the 
coexistence of the two phases was found at certain points in the phase 
diagram.

\begin{figure}[t]
	\subfigure[][]{
		\includegraphics[height=0.197\textwidth]{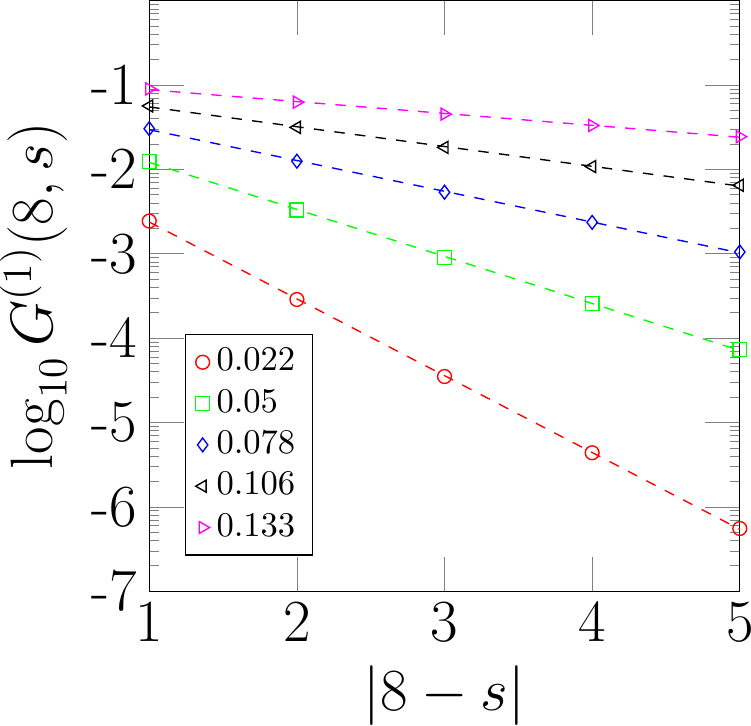}
		\label{fig:JCHG1MI}
	}
	\subfigure[][]{
		\includegraphics[height=0.197\textwidth]{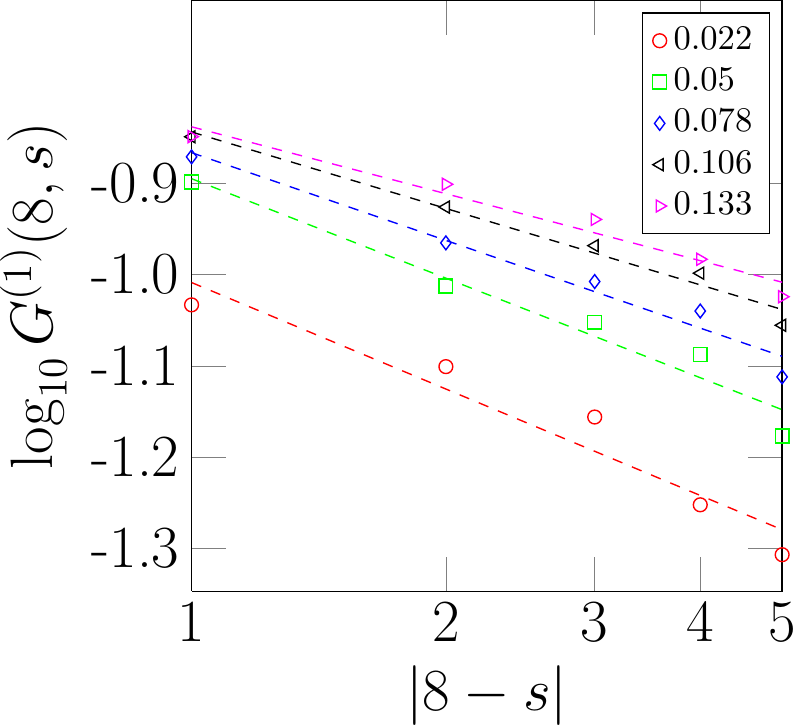}
		\label{fig:JCHG1SF}
	}
\caption{Representative plots of the correlation function $G^{(1)}(8,s)$ for 
the spin excitations in the JCH model with $L=15$. (a)~$n=1$ MI phase at point
$\tilde{\mu} = -0.900$, and (b)~SF phase at point 
$\tilde{\mu} = -0.989$. In both cases the same five values of $\kappa/g$ 
are chosen with values given in the legends. The correlation functions are 
consistent with (a)~exponentials and (b)~power-laws.}
	\label{fig:g1repJCH}
\end{figure}

As discussed in Sec.~\ref{subsec:BH}, in the MI phase one 
expects this correlation function to decrease exponentially with distance, 
$G^{(1)}(r,s)\sim\exp(-|r-s|/\xi)$ over a correlation length $\xi$, while in 
the SF phase it should behave as a power law, 
$G^{(1)}(r,s)\sim 1/|r-s|^{\alpha}$ where $\alpha$ is some positive constant.
The correlation functions are calculated for the JCH model (in which case 
$b\to a$ and $\sigma$ for photons and spins, respectively) and the BH model 
for systems of size $L=15$, 19, 23, 27, and 31. 
Consider for concreteness the shortest length $L=15$, for which the fits are 
the least reliable; the results are shown in Fig.~\ref{fig:g1repJCH}. This size
is chosen convey the worst-quality results for various system sizes. The 
correlations are measured with respect to the central site, in this case site 
number $8$.  The point where $r=s$ is not plotted or used for the fit, since 
only the asymptotic form of the correlations for $|r-s| \gg 0$ is of interest. 
To mitigate boundary effects, the two sites closest to the (open) boundary of 
the system were also not considered. The correlations for the spins and the 
photons track each other very closely, so the results for photons are not 
presented.

The results indicate that the spin excitations and photons behave in close 
analogy to the 
bosons of the BH model; that is, the single-particle correlation functions are
clearly exponential within the MI lobe and follow power laws in the SF regime. 
The power law relationship holds up well not only for $L=15$, but also for all 
larger values of $L$ considered. The exponent associated with the power law, 
corresponding to the slope of the log-log fit, is increasingly precise for 
larger values of $L$, with the standard error $\sim 1/\sqrt{L}$.
The correlation function decreases most rapidly deep in the MI region
but increasingly slowly as the phase boundary is approached. The results shown 
in Fig.~\ref{fig:JCHG1SF} correspond to the SF phase at constant $\mu$, for
values of $\kappa/g$ that range from almost immediately adjacent to the MI lobe
($\kappa/g=0.022$) almost to the edge of the phase diagram ($\kappa/g=0.133$) 
in Fig.~\ref{fig:atomdensities-31}. Unsurprisingly, the power-law fits are poor 
near the phase boundary (c.f.\ the points corresponding to $\kappa/g=0.022$)
but improve as one moves further away. 

The correlations exhibited by each type of carrier also track perfectly with 
each other. This result is in qualitative agreement with Ref.~\cite{Knap2010}, 
in which the same quantity was calculated using the Variational Cluster 
Approximation. In fact, this feature persists for all quantities discussed
below, unless mentioned explicitly. An intriguing consequence of the identical
behavior for the
two species of excitations is that even though one normally views the atoms as 
mediating photonic interactions, one could just as well think of the photons as 
mediating atomic interactions; though the atoms are each isolated within their 
own cavities, they nevertheless feel each others' presence. 

\begin{figure}[t]
	\subfigure[][]{
		\includegraphics[height=0.216\textwidth]{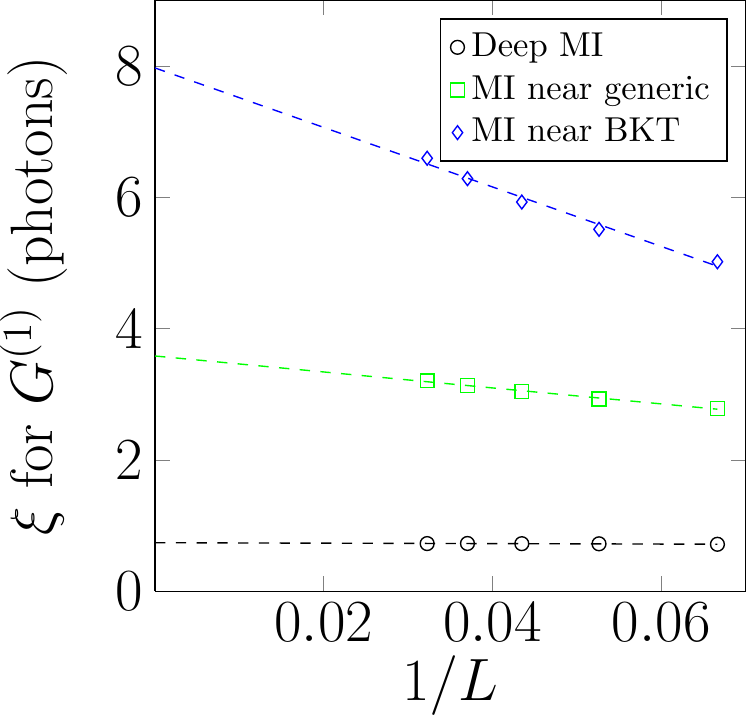}
	}
	\subfigure[][]{
		\includegraphics[height=0.216\textwidth]{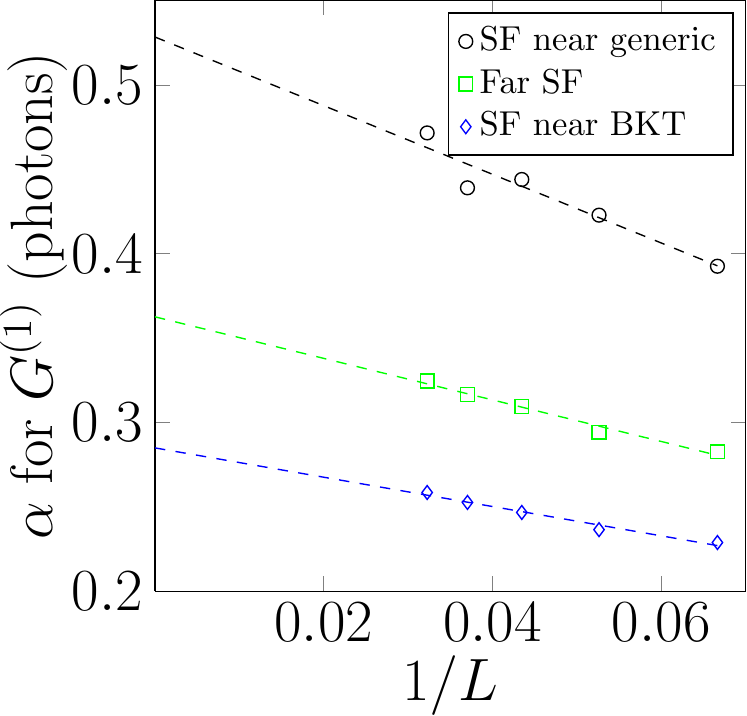}
	\label{fig:scalingSF}
	}
\caption{Representative finite-size scaling analysis of $G^{(1)}$ in the (a) 
MI and (b) SF regimes of the JCH model. The data points for the MI regime are 
$(\tilde{\mu},\kappa/g)=(-0.856,0.044)$ (deep within MI 
lobe), $(-0.856,0.128)$ (near generic transition), and $(-0.900,0.156)$
(near BKT point). The SF regime points are at $(-0.989,0.022)$ (near generic
transition), $(-0.989,0.106)$ (deep within SF phase), and $(-0.989,0.156)$
(near BKT point).}
\label{fig:g1repfss}
\end{figure}

The infinite-system values of the correlation length $\xi$ and the power 
$\alpha$ are estimated using a finite-size scaling analysis, as shown in 
Fig.~\ref{fig:g1repfss} for a few representative points in phase space. The 
best exponential or power-law function is fit for $G^{(1)}$ for each value of 
$L=15$, 19, 23, 27, and 31, such as is shown in Fig.~\ref{fig:g1repJCH} for 
$L=15$. The values of $\xi$ and $\alpha$ are then plotted as a function of 
$1/L$, and the data are fit to a line whose intercept is interpreted as the 
corresponding value in the thermodynamic limit. The data are only weakly 
dependent on system size deep in the MI phase and near the generic phase
boundary, but show a strong dependence near the BKT point. On the SF side the 
values of $\alpha$ are size-dependent for all three phase space points 
considered; this likely reflects the fact that along the $\tilde{\mu}=-0.989$ 
line the Mott boundary remains nearby. The same procedure is carried out for 
the BH model for comparison (not shown).

\begin{figure}[t]
	\subfigure[][]{
		\includegraphics[height=0.170\textwidth]{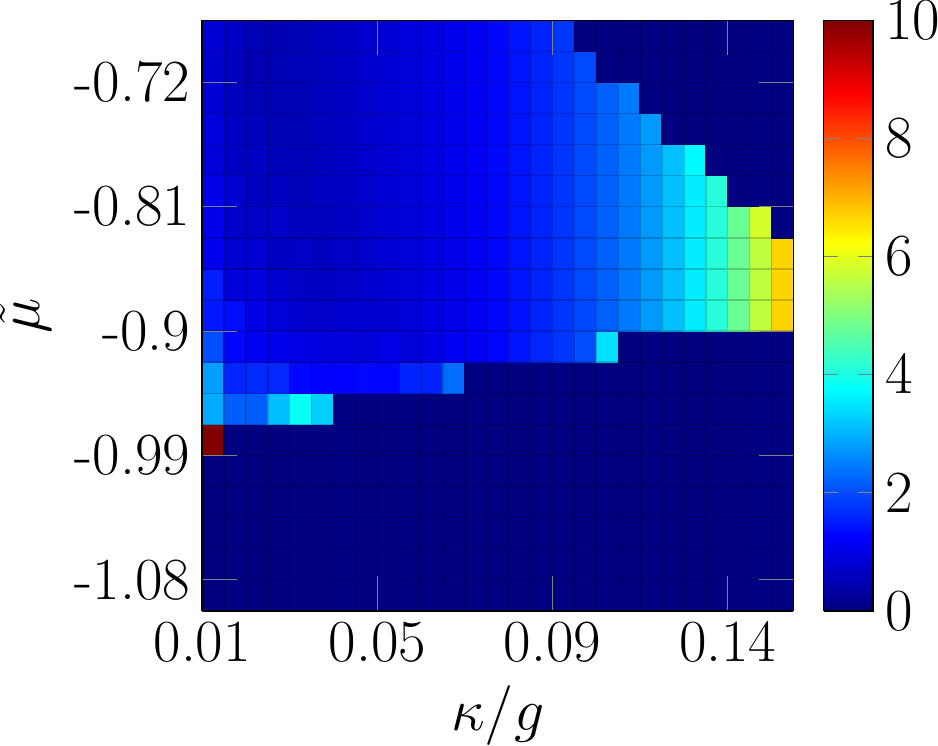}
	}
	\subfigure[][]{
		\includegraphics[height=0.170\textwidth]{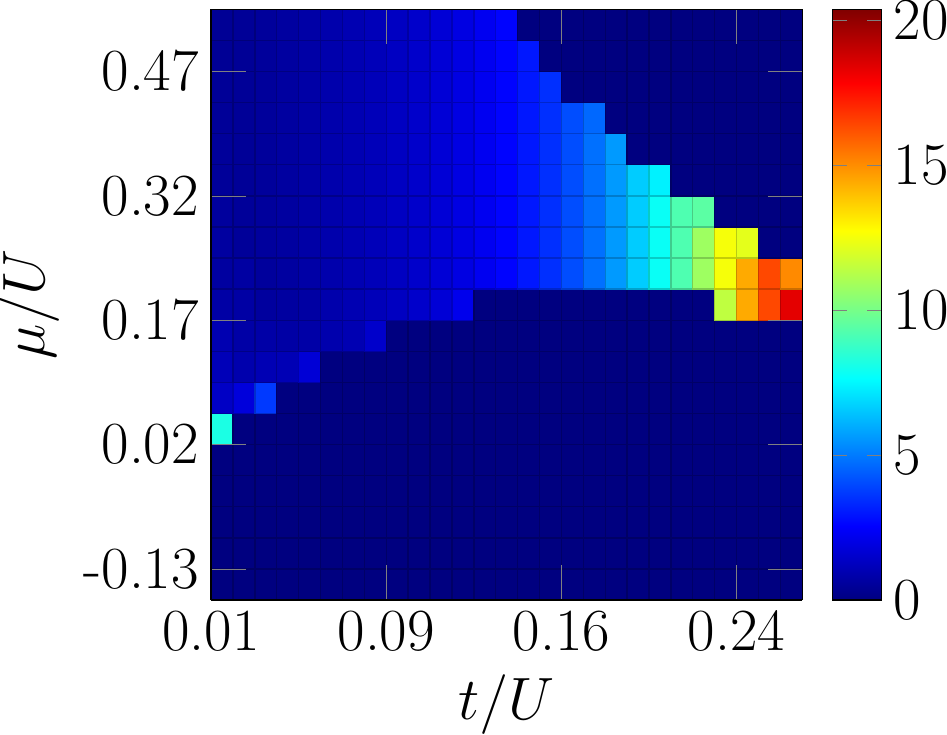}
	}
\caption{Phase diagrams for $\xi$ in the thermodynamic limit, assuming 
$G^{(1)}(r)\sim\exp(-r/\xi)$ in MI regime of the (a) JCH model (photons only) 
and the (b) BH model. The correlation length $\xi$ approaches the system size
at the edge of the Mott lobe near the BKT point.}
	\label{fig:g1xiJCH}
\end{figure}

The phase diagram for the value of $\xi$ in the thermodynamic limit is 
displayed for the JCH and BH models in Fig.~\ref{fig:g1xiJCH}. Note however 
that one cannot obtain $\xi$ in this way immediately at the phase boundary. 
The true correlation length is expected to diverge, and a diverging correlation 
length cannot be accurately captured by a DMRG procedure with finite truncation 
$M$~\cite{Schollwock2011}. Furthermore, the precise location of the phase 
boundary varies depending upon the size of the system, so a particular point 
in phase space near the boundary displayed for $L=31$ may or may not be in the 
Mott-insulating lobe, depending upon the system size. 
In Fig.~\ref{fig:g1xiJCH}, only the phase space points that are unambiguously 
within the $n=1$ lobe have been included; this accounts for the apparently
smaller MI lobes than are depicted in Fig.~\ref{fig:densities}. 

The numerical results clearly indicate that the correlation length is 
independent of $\mu$ and is solely a function of $\kappa$ or $t$, smoothly 
increasing with increasing hopping. Since everywhere within the Mott lobe the 
state has a well-defined number of excitations $N_{\text{tot}}$, the only 
effect of varying the chemical potential by an amount $\Delta\mu$ while fixing 
$\kappa$ or $t$ is to shift the entire spectrum by an amount 
$N_{\rm tot}\Delta\mu$. The ground state itself at fixed hopping is therefore 
independent of $\mu$ within the lobe. The correlation function only increases 
as the BKT point is approached, not near the generic phase boundaries.

\begin{figure}[t]
	\subfigure[][]{
		\includegraphics[height=0.172\textwidth]{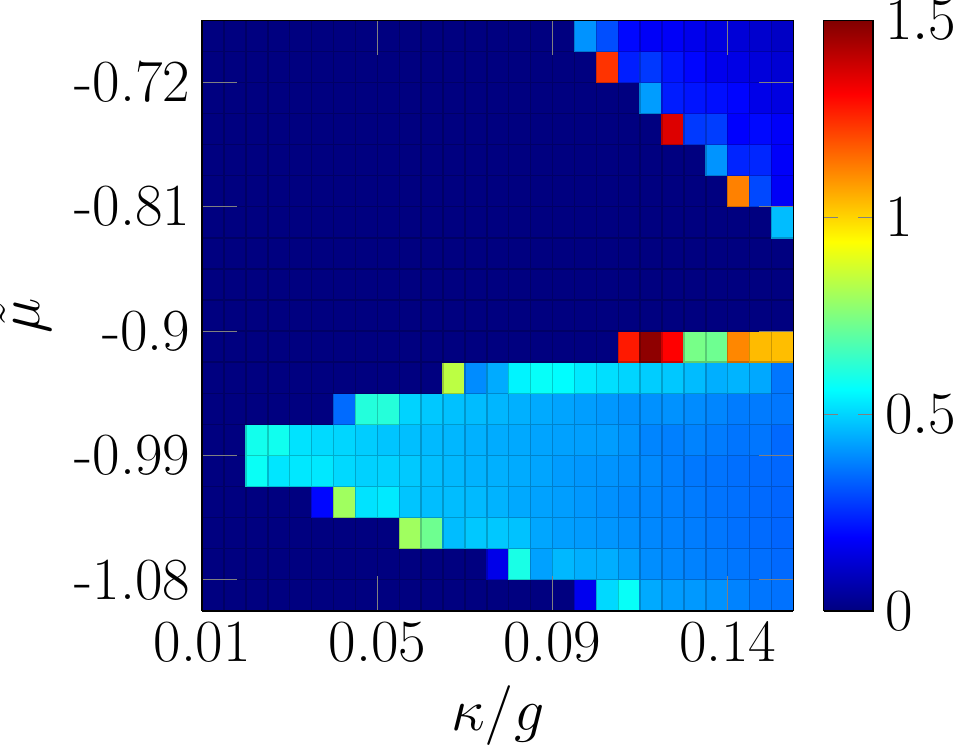}
	}
	\subfigure[][]{
		\includegraphics[height=0.172\textwidth]{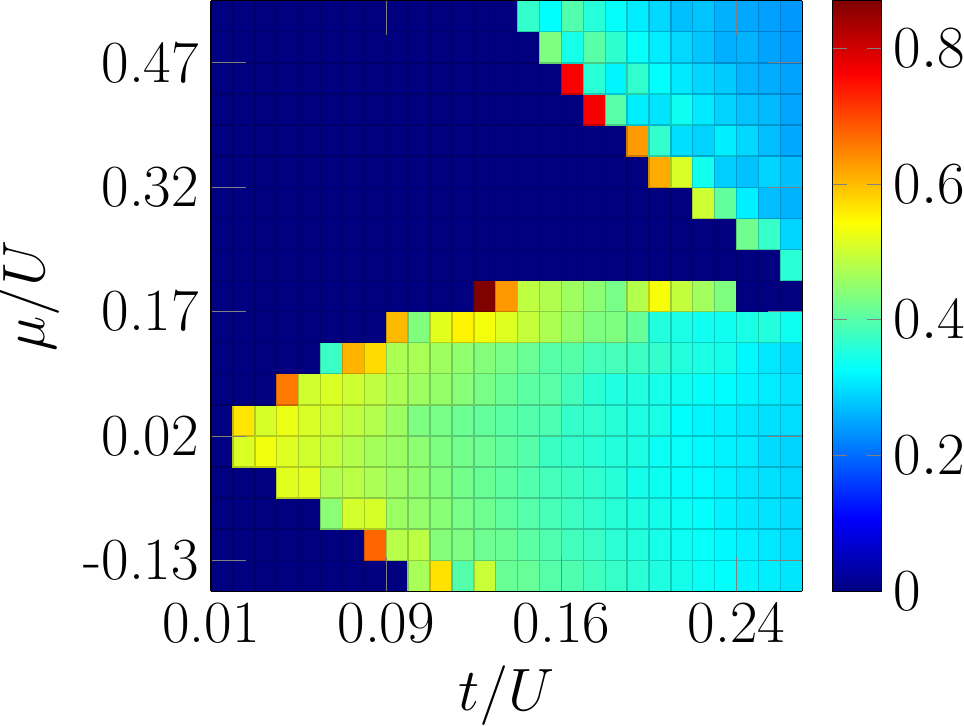}
		\label{BHalpha}
	}
\caption{Phase diagrams for $\alpha$ in the thermodynamic limit, assuming 
$G^{(1)}(r)\sim r^{-\alpha}$ in SF regime of the (a) JCH model (photons only) 
and the (b) BH model. These values of $\alpha$ indicate strong fermionization
of the photons in the JCH model.}

	\label{fig:g1alphaJCH}
\end{figure}

The value of $\alpha$ in the thermodynamic limit is calculated for all points 
in the superfluid phase for both the JCH and BH models, and the results are 
shown in Fig.~\ref{fig:g1alphaJCH}. Right at the phase boundary, the exponent 
approaches values on the order of unity or higher. The same caveats mentioned 
above for the calculation of $\xi$ apply here as well. In addition, the spatial
dependence of $G^{(1)}$ (an example of which is shown in 
Fig.~\ref{fig:JCHG1SF}) and the finite-size scaling data (an example of which
is shown in Fig.~\ref{fig:scalingSF}) are much noisier right near the phase 
boundary. 

Everywhere near the phase transition, however, the value of $\alpha$ 
is close to $0.5$. This value is consistent with what would be expected for a 
Tonks-Girardeau gas of photons [c.f.\ Eq.~(\ref{eq:G1Tonks})] and matches the
Luttinger parameter $K_b=1$, as discussed in Sec.~\ref{subsec:densities}. In 
fact, $\alpha$ has previously been used to obtain the location of the phase 
boundary for the BH model, using infinite system DMRG with periodic boundary
conditions~\cite{Kuehner1998}. This 
indicates that the low-density regime outside of the Mott lobes is in fact not 
strongly superfluid in nature. Rather, the results are consistent with the
complete fermionization of the BH bosons and the JCH photons in the equivalent 
regime. Furthermore, since $G^{(1)}$ for the spin excitations and photons in 
the JCH model 
track each other so well (not shown), the spin excitations have also been 
fermionized, even though the atoms are treated as spins with no particular 
exchange statistics. The `SF' designation of this phase therefore appears to be 
a misnomer, but it will be kept for clarity of exposition in what follows.

The value of $\alpha$ decreases for increasing hopping $\kappa$ or $t$, but the
trend is slow for the parameter range studied. By the edge of the plots in 
Fig.~\ref{fig:g1alphaJCH}, the exponent for the SF phase between the $n=0$ and 
$n=1$ lobes has dropped to almost constant (in terms of $\mu$) values of 
$0.33$ and $0.29$ for the JCH and BH models, respectively. For larger values of
$\mu$ where the density is higher the exponent drops off more rapidly, 
reaching a range of 0.081-0.254 for the JCH model and 0.226-0.336 for the BH 
model at the edge of the plots. These values are all quite different from the 
value $\alpha=0$ that one would expect for an ordinary superfluid, however.

\subsubsection{Two-body correlation function}

One of the most important signatures of fermionization within the low-density 
SF phase is found in the two-body correlation function $G^{(2)}(r,s)$ defined 
in Eq.~(\ref{eq:G2}) and its normalized variant $g^{(2)}(r,s)$ defined in 
Eq.~(\ref{eq:g2Tonks}), as discussed in Sec.~\ref{subsec:Tonks}. Within the 
$n=1$ Mott-insulating lobe, one expects 
$g^{(2)}(r,s)=1-\delta_{r,s}$ irrespective of the model and 
the excitation, which is exactly what is observed. Of greater interest is the 
behavior of this correlation function in the SF regime for different mean
densities. For a perfect superfluid the two-body correlation function is 
featureless, $g^{(2)}(r,s)=1$ in the bulk, reflecting the fact that all 
superfluid carriers occupy the same plane wave state. On the other hand, for a 
system of free fermions, the two-body correlation function exhibits two 
important features. The first is the presence of an exclusion hole at $r=s$ 
reflecting the Pauli exclusion 
principle. The second is the characteristic Friedel oscillations appearing on 
either side of the exclusion hole, with a wavelength $\lambda_F$ set by the
mean density $n$ or Fermi wavelength $k_F$, as discussed in Sec.~\ref{subsec:Tonks}. 

\begin{figure}[t]
	\subfigure[][]{
		\includegraphics[height=0.200\textwidth]{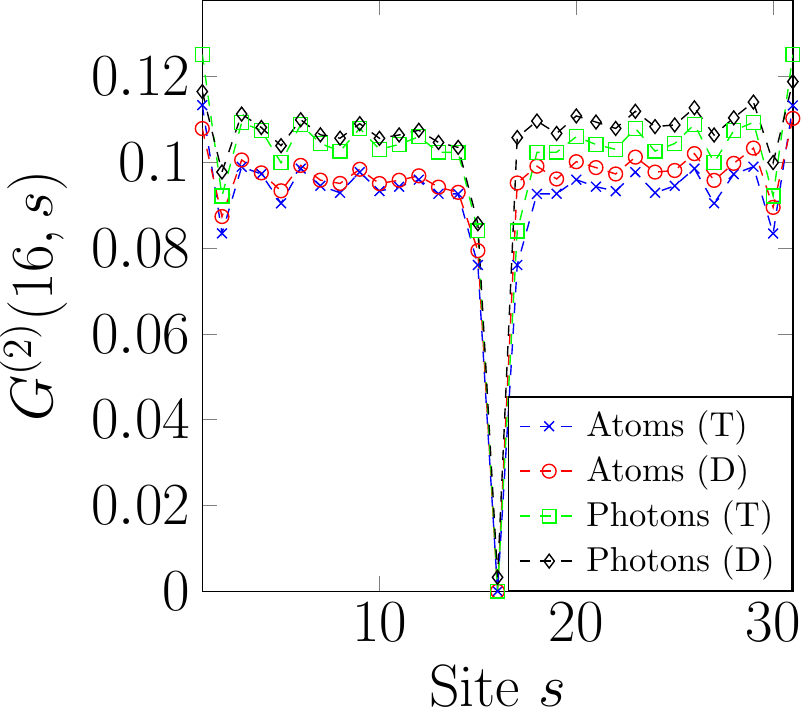}
	}
	\subfigure[][]{
		\includegraphics[height=0.200\textwidth]{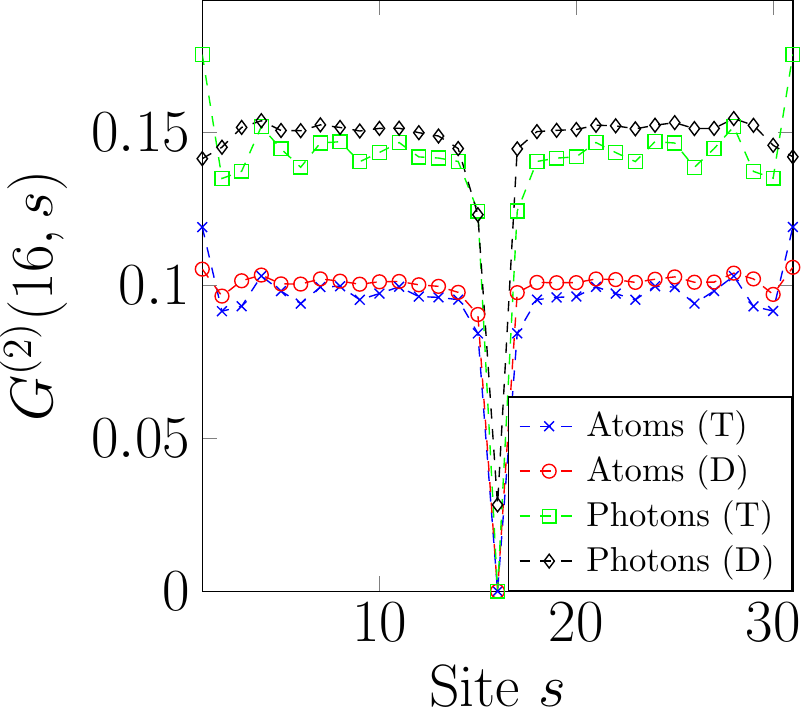}
		\label{fig:G2JCH2}
	}\\
	\subfigure[][]{
		\includegraphics[height=0.200\textwidth]{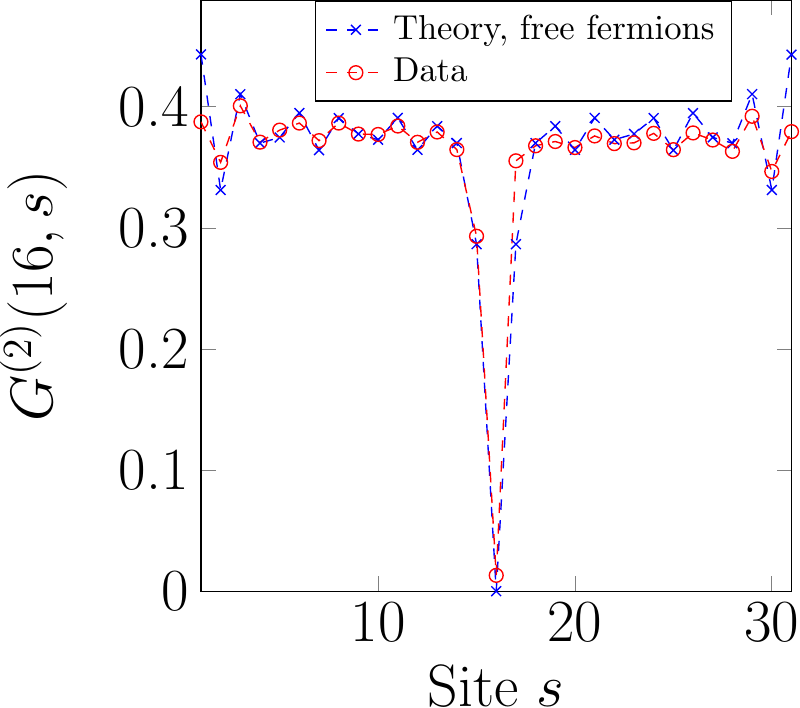}
	}
	\subfigure[][]{
		\includegraphics[height=0.200\textwidth]{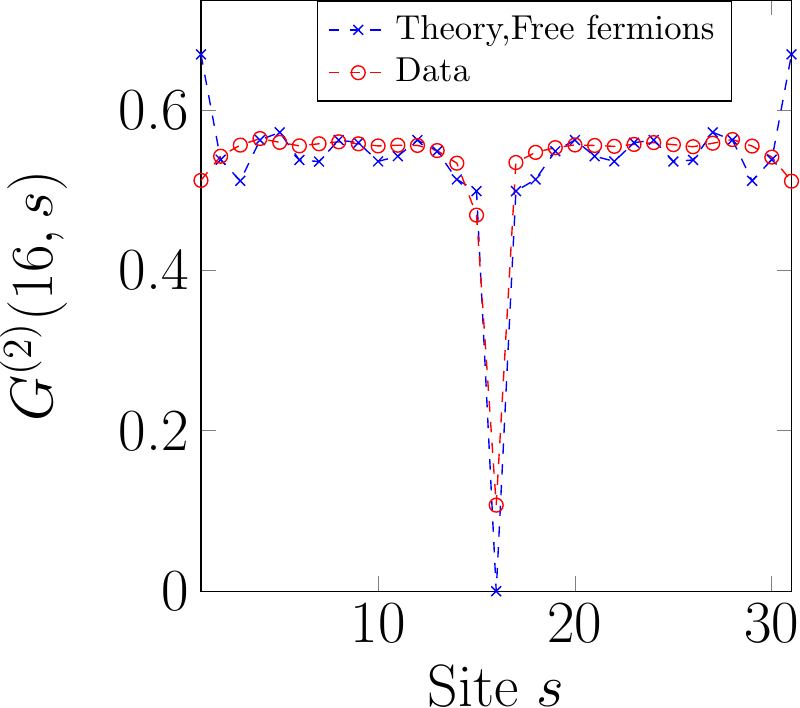}
		\label{fig:G2BH2}
	}\\
\caption{The spatial-dependence of the unnormalized two-body correlation 
function $G^{(2)}(16,s)$ for $L=31$ is shown for representative points in the 
SF phase where $\alpha \approx \frac{1}{2}$. Figures (a) and (b) correspond to 
photons at the points $(\tilde{\mu},\kappa/g)=(-0.989,0.039)$ and 
$(-0.989,0.106)$ in the JCH model, respectively; `T' and `D' in turn denote 
`theory' and `data.' Figures (c) and (d) correspond to bosons at the points 
$(\mu/U,t/U)=(0.019,0.066)$ and $(0.019,0.180)$, respectively.}
	\label{fig:g2BH}
\end{figure}

The unnormalized two-body correlation functions $G^{(2)}(r,s)$ for the photons
and spin excitations in the JCH model are plotted in Fig.~\ref{fig:g2BH} for 
representative points in the SF phase where $\alpha\approx\frac{1}{2}$; the BH 
results are also shown for comparison. (The momentum distribution $n(k)$ in the
1D BH model was previously considered for various interaction strengths within 
the $n=1$ MI lobe~\cite{Ejima2011}, and for both the MI and SF regimes at unit 
filling~\cite{Kollath2004}). The data are compared with the analytical 
expression for free fermions at the same mean density, Eq.~(\ref{eq:g2FF}). The 
unnormalized two-body correlation function is plotted
in order to make apparent the particle densities for the particular points in 
phase space. The two main signatures of fermionization in the SF phase, the 
exclusion hole and the Friedel oscillations, are evident in all the plots. At
lower densities, the match between the data and the prediction based on 
non-interacting fermions is excellent. This indicates that the superfluid
density at this point in the SF phase is close (if not exactly equal) to zero. 
One would expect the systems to behave more like a superfluid as the tunneling 
is increased and the excitation density increases, for points in phase space
that are further from the phase boundary. Indeed this is the case; the data 
depicted in Figs.~\ref{fig:G2JCH2} and \ref{fig:G2BH2} show that the exclusion 
hole is now slightly filled in, and the Friedel oscillations are increasingly 
washed out. 

An interesting feature in the case of the JCH model, not present in the BH 
model, is that the wavelength of the oscillations for both the photons and the 
bosons is set by the mean density of polaritons. The amplitude of $G^{(2)}$ for 
each carrier is determined by the density of just that carrier, however. More 
precisely, to obtain a fitting curve of the form of Eq.~(\ref{eq:g2FF}) for 
$G^{(2)}(r,s)$ for the photons (spin excitations), one must use the mean number
of photons (spin excitations) as the value of $N$ in the prefactor 
$\left(N/(L+1)\right)^2$, but the number of polaritons in the oscillatory
functions appearing in $A(r,s)$ and $B(r,s)$. Hence, the spin excitations and 
the photons are individually fermionized, inasmuch as they have inherited this 
property from the polaritons.

\subsection{Other measures of the ground state}
\label{subsec:others}

\subsubsection{Superfluid fraction}

The most compelling evidence for the existence of a SF phase would be the
presence of a non-zero superfluid order parameter. An example of such an order 
parameter is the superfluid fraction or superfluid stiffness $f_s$, the ratio 
of particles exhibiting superfluid flow to the total number of particles. The 
factor $f_s$ can be calculated numerically by imposing periodic boundary 
conditions and applying a phase twist $\Theta \ll \pi$ to the boundary 
conditions~\cite{Fisher1973,Krauth1991,Singh1994,Roth2003}. In practise, this 
can be accomplished by means of a Peierls factor applied to the bosonic 
creation and annihilation operators in the BH and JCH models, which has the 
effect of modifying the hopping terms via $a_i^{\dagger} a_j \longmapsto 
a_i^{\dagger} a_j e^{-\mathrm{i} \Theta/L}$~\cite{Roth2003a}. While this 
induces a velocity $v=(2J)\nabla\Theta$ for each quantum particle ($J$ 
is the hopping coefficient corresponding to $\kappa$ in the JCH model or $t$ 
in the BH model) because the current density is $j=nv$, only the particles in 
the superfluid will respond collectively. As a result, the ground state energy 
will increase relative to the twist-free case solely due to the kinetic energy 
of the superfluid particles. From this change of the ground state energy, the 
superfluid fraction can be determined:
\begin{equation}
f_s = \frac{L}{2Jn}\left.\frac{\partial^2 E_0(\Theta)}{\partial\Theta^2}
\right|_{\Theta=0}
\approx\frac{L}{Jn} \frac{E_0(\Theta)-E_0}{\Theta^2},
\label{eq:sffracformula} 
\end{equation}
where $E_0$ is the ground state energy with no phase twist and $E_0(\Theta)$ 
is ground state energy with overall twist $\Theta\ll\pi$. The latter expression
is a finite-difference approximation for the second derivative of the energy 
with respect to the phase twist, using the central three-point stencil. This 
calculation of $\rho_s$ has previously been performed at constant density near 
the BKT transition of the 1D BH model across the tip of the $n=1$ Mott 
lobe~\cite{Pai1996}; a significant jump in $\rho_s$ across the transition was 
found in that work. This work instead examines $\rho_s$ at representative 
points in the low-density SF phase far from the transition, with considerably 
different results.

In the numerical calculations, we considered various points in the SF phase
using five different finite-size systems $L=15,19,23,27,31$. Periodic boundary 
conditions are required, which are computationally more demanding for the DMRG 
method than are open boundary conditions (see the discussion in 
Sec.~\ref{subsec:methods}); hence, only a set of representative points in the 
SF phase were considered. Recall from Sec.~\ref{subsec:methods} that the $n=1$ 
BKT points in the JCH and BH models are located at approximately 
$(\tilde{\mu},\kappa/g)=(-0.95,0.2)$~\cite{Rossini2007,Mering2009} and
$(\mu/U,t/U)=(0.09,0.3)$~\cite{Kuehner1998,Kuehner2000,Ejima2011}, 
respectively. We therefore considered these three points in the SF phase of the 
JCH model: $(-0.944, 0.133)$, $(-0.989, 0,028)$, and $(-1.00, 0.240)$. The 
first is just left of the BKT point, in the vicinity of the hole boundary, the 
second is between the $n=0$ and $n=1$ Mott lobes, and the third is to the right 
of the BKT point. In particular, the first two points correspond to mean polariton densities $n < 1$, while the third has $n >1$. For the BH model we considered the two points: $(0,0.08)$ and 
$(0,0.5)$; again, the first is left of the BKT point in the vicinity of the 
hole boundary, while the second point is much to the right, in the deep SF 
phase well beyond the region depicted in Figs.~\ref{BH-density-31} and 
\ref{BHalpha}. The first of these points has mean boson density $n < 1$ and the second, $n > 1$. 

The ground-state energy was obtained for these points in the SF region, for 
phases $\Theta/\pi\in(0.0, 1.0)$ in increments of $0.2$. Note that the ground 
state 
energy is invariant under the transformation $\Theta \longmapsto -\Theta$. For 
each value of $L$, the second derivative in Eq.~(\ref{eq:sffracformula}) was 
then calculated using $\Theta=0.2\pi$. The results were then verified by 
estimating the derivative using a central-difference approximation with 
five-point ($\Theta/\pi \in \set{0,\pm 0.2,\pm 0.4}$) and seven-point stencils 
($\Theta/\pi \in \set{0,\pm 0.2,\pm 0.4,\pm 0.6}$); no discernible difference 
from the method of Eq.~(\ref{eq:sffracformula}) was observed in the results so 
obtained. The values of $f_s$ were also obtained by extracting the coefficient 
for the quadratic term in a polynomial fit of $E_0(\Theta)$. Again, no 
discernible differences from the central-difference results were found using 
this approach. Once the value of $f_s$ was obtained for a given system size, a 
finite-size scaling analysis in $1/L$ was performed to interpolate to the 
thermodynamic limit. As a final check on the results, the calculation in 
Eq.~(\ref{eq:sffracformula}) was repeated with smaller values of the phase 
twist, $\Theta/\pi\in[0.0,0.2]$ in increments of $0.02$. For 
$\Theta/\pi < 0.06\pi$, the values of $\rho_s(L)$ could not be reliably fitted 
to determine the value in the thermodynamic limit (the problem of dividing one 
small number by another). However, for $\Theta/\pi \geq 0.06$, the results were 
consistent with those obtained with $\Theta/\pi=0.2$.

\begin{figure}[t]
	\subfigure[][]{
		\includegraphics[height=0.236\textwidth]{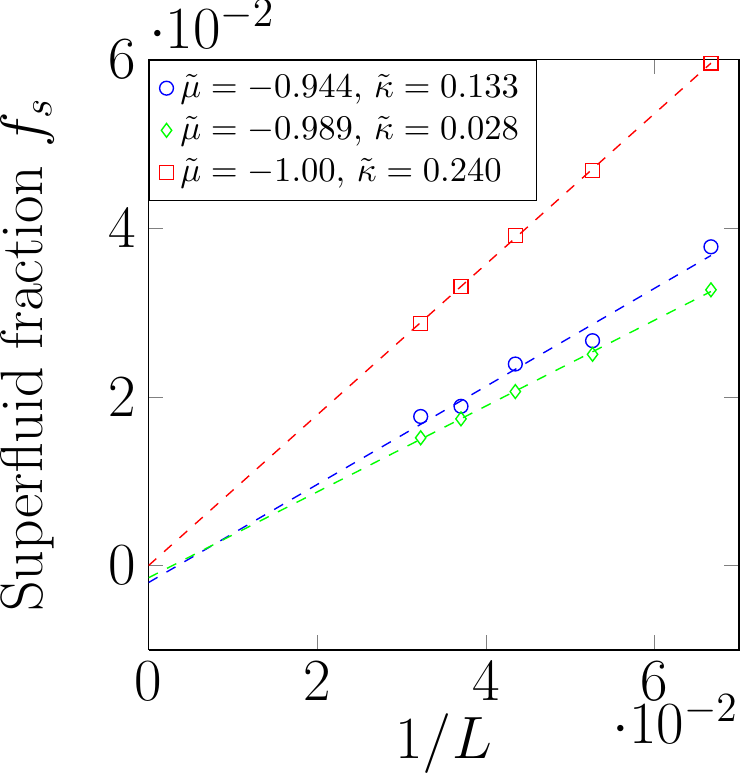}
	}
	\subfigure[][]{
		\includegraphics[height=0.236\textwidth]{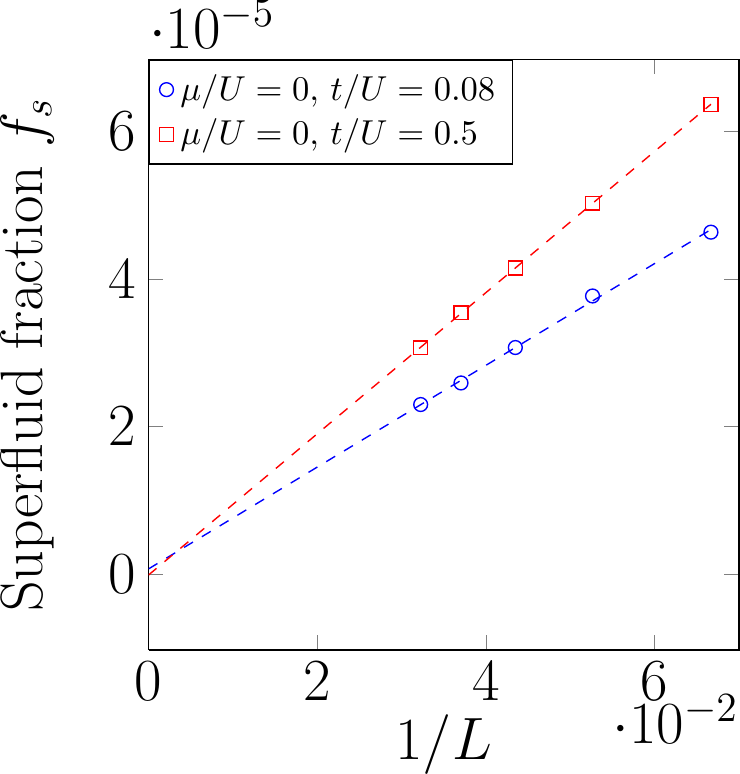}
		\label{fsBH}
	}
\caption{Superfluid fraction for the (a) JCH and (b) BH model outside but 
near the $n=1$ MI lobe. For the JCH model, the symbol $\tilde{\kappa}=\kappa/g$
is used for 
compactness. For both systems, the curve marked by red squares (color online) 
is beyond the critical hopping value for the BKT transition for the $n=1$ MI 
lobe, while the others are not. In both cases, the superfluid fraction $f_s$ 
tends to 0 in the thermodynamic limit.}
	\label{fig:SFfraction}
\end{figure}

\begin{table}[t]
\centering
\begin{tabular}{|cccc|cccc|}
  \hline
  \multicolumn{4}{c}{BH model}  & \multicolumn{4}{c}{JCH model} \\
  \hline
  $\mu/U$ & $t/U$ & $f_s$ & $\Delta f_s$ &  $\tilde{\mu}$ & $\kappa/g$ & $f_s$ & $\Delta f_s$ \\
  \hline
  0 & 0.08 & 0.0007 & 0.0008 &  -0.944 & 0.133 & -0.0020 & 0.0025  \\
  0 & 0.5 & -0.0001 & 0.0001 &  -0.989 & 0.028 & -0.0014 & 0.0004  \\
  & & & &  -1.00 & 0.240 & 0.0000 & 0.0004  \\
  \hline
\end{tabular}
\caption{Numerical values (with uncertainties) of superfluid fraction $f_s$ in thermodynamic limit, for phase space points indicated in Fig.~\ref{fig:SFfraction}.}
\label{tab:fsTDL}
\end{table}

The finite-size scaling results for the superfluid fraction $f_s$ are shown in 
Fig.~\ref{fig:SFfraction}, and the resulting values and uncertainties of $f_s$ 
in the thermodynamic limit are displayed in Table~\ref{tab:fsTDL}. These values 
are determined by the standard least-squares minimization procedure for linear 
fitting. In all cases, except for the JCH point marked by red squares in 
Fig.~\ref{fig:SFfraction}, the obtained thermodynamic limit value of the 
superfluid density contains 0 within its error interval. For the one 
exceptional case, the thermodynamic limit $f_s$ is an order of magnitude 
smaller than the finite-size values (as well as unphysically negative), 
indicating that the superfluid fraction should vanish in the thermodynamic 
limit. It is interesting to note that the finite-size scaling plot for the BH 
point $(\mu/U,t/U)=(0.5,0.5)$ (not shown), which is situated above the particle 
boundary of the $n=1$ lobe, is identical to the scaling of the $(0.0, 0.5)$ 
point shown in Fig.~\ref{fsBH}. Similar behavior is found in the JCH model in
the SF region to the right of the BKT point. Consider for example the
point $(\tilde{\mu},\kappa/g)=(-0.750, 0.240)$ located above the 
$(-1.000, 0.240)$ point in Table~\ref{tab:fsTDL}. The infinite-system
superfluid fraction inferred from finite-size scaling is 
$f_s=0.0006\pm 0.0033$, consistent with zero. This indicates that the 
superfluid density is only weakly dependent on the chemical potential for a 
given hopping strength. The data strongly suggest that the superfluid density 
is zero throughout the low-density SF region studied.

\subsubsection{Condensate fraction}

The condensate fraction $f_c$ is defined as the proportion of particles in the 
lowest-lying single-particle eigenstate of the system. In practice, this can
be obtained by the largest eigenvalue of the single-particle density 
matrix $G^{(1)}(r,r^{\prime})$~\cite{Penrose1956,Penrose1951}.
In fact the entire spectrum, known as the entanglement spectrum~\cite{Li2008b},
can be used to identify quantum phases. In the atomic limit, the 
single-particle ground state for a system of length $L$ is $L$-fold degenerate,
since there is no preferred lattice site. In the non-interacting limit, the
finite-size systems have a sinusoidal single-particle ground state. At zero 
temperature, one expects macroscopic occupation of the ground state in the SF
regime, tending towards unit occupation for large $J/U$. In the MI regime one 
expects the particles to be 
distributed evenly across many nearly-degenerate single particle states, a 
phenomenon known as fragmentation for large occupation of a single 
site~\cite{Spekkens1999,Mueller2006}. If the particles have fermionized, 
however, then no 
such macroscopic occupation should occur; $f_c$ can therefore be viewed as 
another signature of fermionization.

\begin{figure}[t]
	\subfigure[]{
		\includegraphics[height=0.216\textwidth]{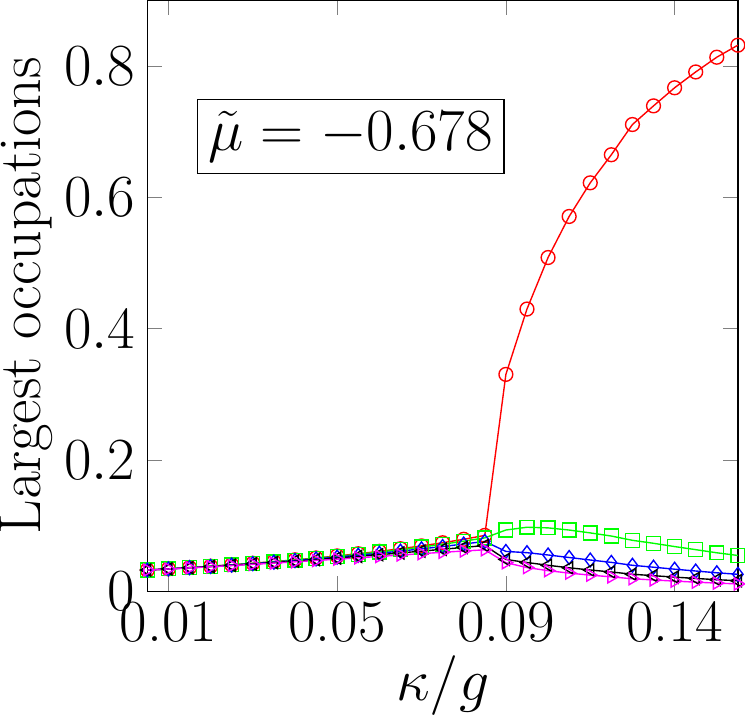}
	}
	\subfigure[]{
		\includegraphics[height=0.216\textwidth]{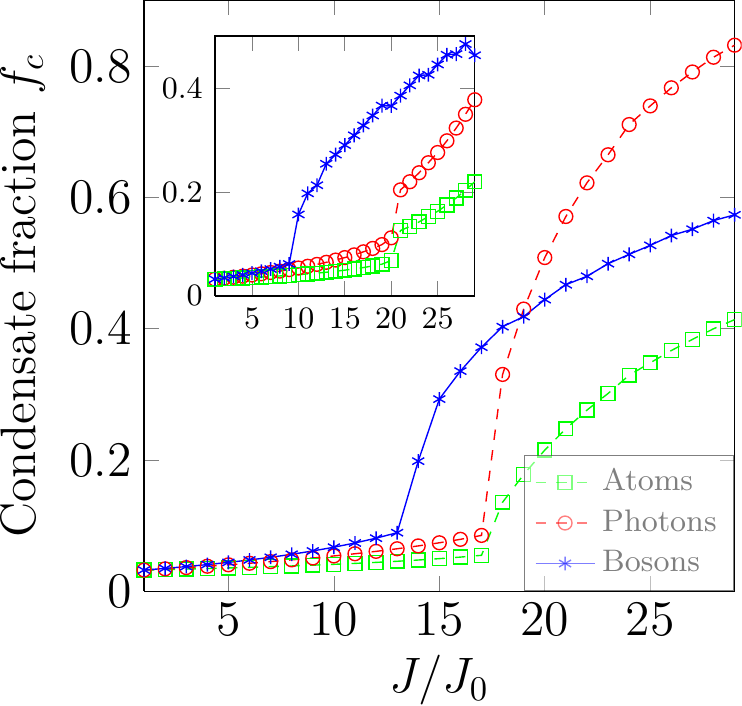}
	}
\caption{Maximal eigenvalues of the reduced single-particle density matrix for
$L=31$. In (a), the five largest eigenvalues of the photon density matrix are 
shown for fixed $\tilde{\mu}=-0.678$ as a function of hopping amplitude 
$\kappa/g$. In (b), the largest eigenvalue of the density matrix (identified 
with the condensate fraction $f_c$ in the SF phase) for spin excitations and 
photons in the JCH 
model, and for bosons in the BH model, are shown as a function of generalized 
hopping $J/J_0$ at fixed chemical potential $\tilde{\mu}=-0.678$
and $\mu/U=0.55$, respectively. The inset shows the lower density case
$\tilde{\mu}=-0.922$ and $\mu/U=0.133$ for comparison.}
	\label{fig:BECfraction2}
\end{figure}

The five largest eigenvalues of the single-particle photon density matrix are 
plotted in Fig.~\ref{fig:BECfraction2}(a) in the JCH model for 
$\tilde{\mu}=-0.678$ and $L=31$ as a function of hopping strength $\kappa/g$.
This corresponds to the constant $\tilde{\mu}$ line near the very top of 
Fig.~\ref{fig:densities}(a). In the zero-hopping limit 
$\kappa/g\to 0$, each photon is perfectly localized to each site, and the 
eigenvalues are precisely $1/L$ (note that the density matrix is normalized to 
unity rather than the total number of particles). As $\kappa/g$ increases the
degeneracy is broken and the largest eigenvalues increase due to the 
fluctuations of the site occupations, while others decrease to preserve the 
normalization (not shown). At a critical hopping strength 
$\kappa/g\approx 0.085$ coinciding with the superfluid transition at this value
of $\tilde{\mu}$, one of the eigenvalues increases precipitously relative to 
the others, signifying the macroscopic occupation of a single mode.
This eigenvalue is associated with the (quasi)condensate fraction. For larger 
hopping strengths the condensate fraction tends towards unity, as expected for 
a non-interacting Bose gas at zero temperature.

Fig.~\ref{fig:BECfraction2}(b) compares the values of the condensate fraction 
for the BH bosons with the JCH photons and spins along the same line of 
constant 
chemical potential considered above, corresponding to $\mu/U=0.55$ in the BH 
model (recall that $\mu/U\sim 1.707(\tilde{\mu}+1)$ as discussed in 
Sec.~\ref{subsec:BH}). The value of $f_c$ for each model is plotted for fixed
size $L=31$ as a function of the hopping $J$, where $J=\kappa$ ($t$)
in the JCH (BH) case, in units of the minimum hopping considered $J_0$ 
corresponding to $\kappa_0/g=0.00556$ and $t_0/U=1.7\kappa/g=0.00948$. Though 
the onset of superfluidity occurs for smaller $J/J_0$ in the 
BH case, consistent with Fig.~\ref{fig:densities}, the condensate fraction does 
not increase as quickly as that of the photons in the JCH case. This might 
simply reflect the fact that the mean carrier density is lower for the BH 
model at the top right point in the phase diagram than for the JCH model, which
would discourage condensation. At lower mean particle densities on the SF side
$\tilde{\mu}=-0.922$ ($\mu/U=0.133$), shown in the inset of 
Fig.~\ref{fig:BECfraction2}(b), the $f_c$ for the BH case is larger than that 
for the photons of the JCH model, but neither reach 50\%. The finite-size
results suggest that $f_c$ tracks the mean particle density, as is discussed
further below.

Interestingly, the spin excitations in the JCH model also show strong 
evidence of condensation. The value of $f_c$ on the SF side reaches 
approximately half that for the photons for the largest value of $J/J_0$
considered for this value of $\tilde{\mu}$. For lower values of $\tilde{\mu}$ 
in the vicinity of the $n=0$ Mott lobe, the ratio of $f_c$
for spin excitations to photons approaches unity (not shown). These results are
generally consistent with observations above that indicate that the spin and 
photon degrees of freedom follow each other closely. The condensation in the 
spin sector therefore appears to be driven sympathetically by the photons via 
the polariton excitations. The results are nevertheless somewhat surprising,
suggesting that the spin excitations are delocalized. 

To obtain the condensate fraction in the thermodynamic limit, the values of 
$f_c$ were obtained throughout the phase diagram for each triple $(\mu, J, L)$, 
where $L=15,19,23,27,31$ using DMRG subject to open boundary conditions.
The finite-size scaling analysis was performed for the SF regime only because 
the procedure is not robust for the points in the MI regime. Nevertheless, 
the finite-size results within the MI regime clearly indicate fragmentation, 
and since the macroscopic degeneracy in the atomic limit is true independent of 
system size, there is no reason to believe that the results in the 
thermodynamic limit would be qualitatively different. 

\begin{figure}[t]
	\subfigure[]{
		\includegraphics[height=0.176\textwidth]{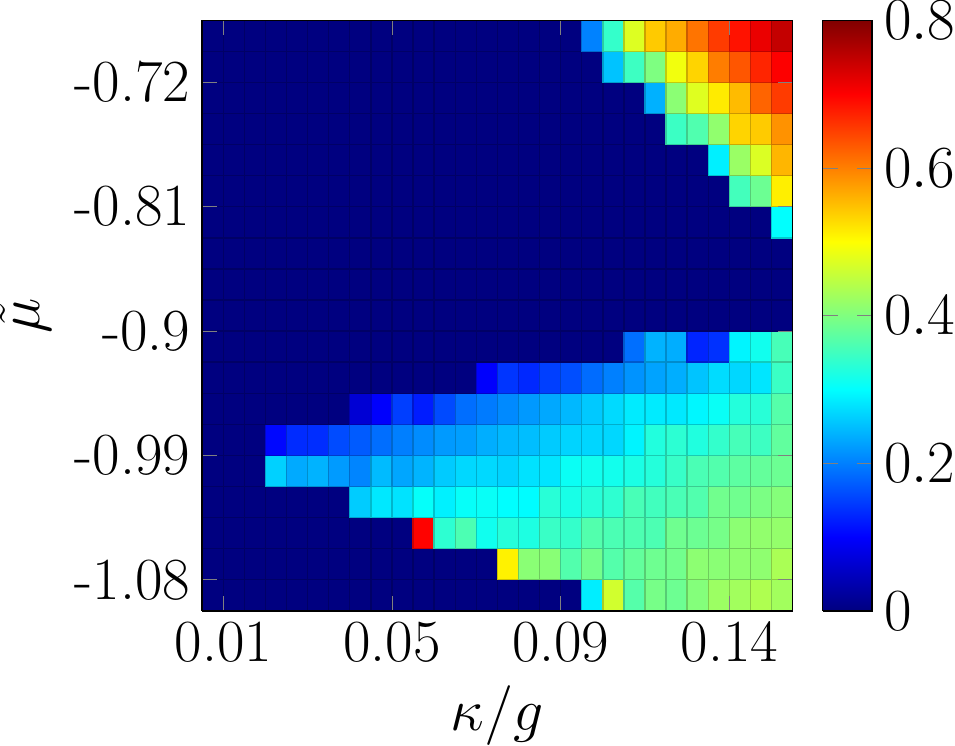}
	}
	\subfigure[]{
		\includegraphics[height=0.176\textwidth]{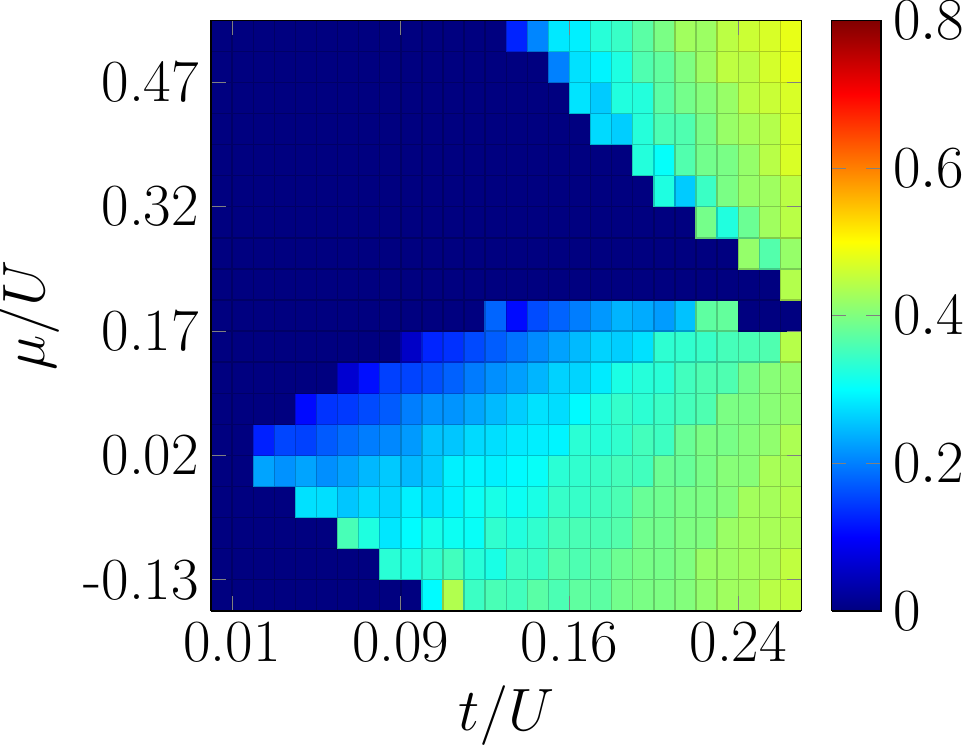}
	}
\caption{Condensate fraction for the (a) JCH model (photons only) and the 
(b) BH model outside but near $n=1$ MI lobe, in the thermodynamic limit.
In both cases, the condensate fraction within the MI lobes is zeroed out.}
	\label{fig:BECfraction}
\end{figure}

The condensate fractions for JCH and BH models in the thermodynamic limit
throughout the explored SF regime of the phase diagram are shown in 
Fig.~\ref{fig:BECfraction}(a) and (b), with the MI lobes explicitly zeroed out. 
Both pictures have the same color scale and are plotted with the axes 
corresponding to equivalent energy scales. The results are qualitatively 
similar for both models, but there are some quantitative differences. The 
value of $f_c$ rises more rapidly with increasing hopping in the JCH model 
than in the BH model, attaining $f_c=0.800$ at the highest mean densities 
investigated, as opposed to only $f_c=0.500$ for the BH model. 
The values of $f_c$ shown in the phase diagrams strongly
resemble the mean densities of spins, photons, and bosons, such shown in 
Fig.~\ref{fig:densities}. Deep in the SF regime the $f_c$ closely follow the 
mean excitation densities, suggesting that the proclivity toward condensation
is governed by the mean density. That said, as the Mott lobe boundary 
approaches the ratios of $f_c$ to the respective mean densities is found to 
increase markedly even as $f_c$ approaches zero.

While there need be no direct relationship between Bose-Einstein condensation
and superfluidity, it is nevertheless somewhat surprising that the condensate 
fraction in the thermodynamic limit would be so large throughout the SF region 
where the superfluid fraction remains zero. These results are nevertheless 
consistent with the small values of $\alpha$ in the SF region, shown in 
Fig.~\ref{fig:g1alphaJCH} (recall that $\alpha\to 0$ for Bose-Einstein 
condensates), as well as with the filling in of the exclusion hole and the 
disappearance of the Friedel oscillations found in the spatial-dependence of 
the two-body correlation function (c.f.\ Fig.~\ref{fig:g2BH}). In addition,
the value of $f_c$ is consistently small in the low-density SF regime, as
expected for a fermionized gas.

\subsubsection{Entanglement properties}
\label{subsubsec:entanglement}

Much can be learned about the ground states of physical systems by examining 
the properties of subsystems. A notable example is the entropy of 
entanglement~\cite{Nielsen2000}. This is obtained by partitioning the system
into a block of contiguous lattice sites $b \subset \mathcal{L}$, where 
$\mathcal{L}$ denotes the full lattice, and its complement $b^{\prime}$. The 
entropy of entanglement associated with this bipartition is given by
\begin{equation}
S_{\rho_b} = -\text{Tr}(\rho_{b} \text{ln}\rho_{b}),
\end{equation}
where $\rho_{b}$ is the reduced density matrix of the state over $b$. 
Analytical formulas for the entanglement entropy of non-interacting fermions
and bosons on a lattice exist for the semi-infinite chain~\cite{Peschel2009} 
but we are not aware of any for finite-size systems greater than a few sites.

The scaling of the entanglement entropy with the size of the subsystem is 
intimately linked with the utility of DMRG as a simulation method. For a wide 
array of physical systems, the entanglement entropy obeys an area 
law~\cite{Eisert2010}, meaning that the entanglement entropy $S_{\rho}$ 
associated with $\rho_b$ is proportional to the number of sites at the 
interface between $b$ and $b'$ (henceforth denoted $l$), rather than to its 
volume or cardinality. In one dimension, the area law corresponds to a value of 
$S_{\rho}$ that saturates for some finite value of $l$:
\begin{equation}
S_{\rho}(l) \leq D,
\label{eq:arealaw}
\end{equation}
where $D$ is a constant, independent of $l$ (of course $l=2$ is itself 
constant). Only states satisfying Eq.~(\ref{eq:arealaw}) can be efficiently 
simulated by DMRG for large system sizes, because the variational ansatz used 
by the DMRG algorithm explicitly assumes that the area law is 
satisfied~\cite{Vidal2003,Eisert2010}. 
Asymptotic scaling results~\cite{Wolf2006} reveal that free fermions always 
logarithmically violate the area law in any dimension in both the continuum 
and on a lattice, whereas free bosons in 1D satisfy the area law away from 
criticality~\cite{Plenio2005}. 

The entanglement entropy associated with a finite block can be used to 
distinguish bosonic and fermionic behavior. The entanglement entropy of the 1D 
non-interacting Bose gas is a smooth function, whereas that
for the non-interacting Fermi gas oscillates with $l$. This is because at zero 
temperature the bosons condense into the smooth lowest-lying eigenstate of the 
hopping model, while the Pauli exclusion principle forces fermions into 
oscillatory excited states. Generally, the entanglement entropy for bosons
is larger than for fermions because the number of accessible states $\Omega_b$ 
is exponentially greater than $\Omega_f$, meaning that when the system is bipartitioned, for each configuration of the left half of the chain there are an exponentially larger number of compatible configurations for the right half in the bosonic case.

\begin{figure}[t]
	\subfigure[][]{
		\includegraphics[height=0.200\textwidth]{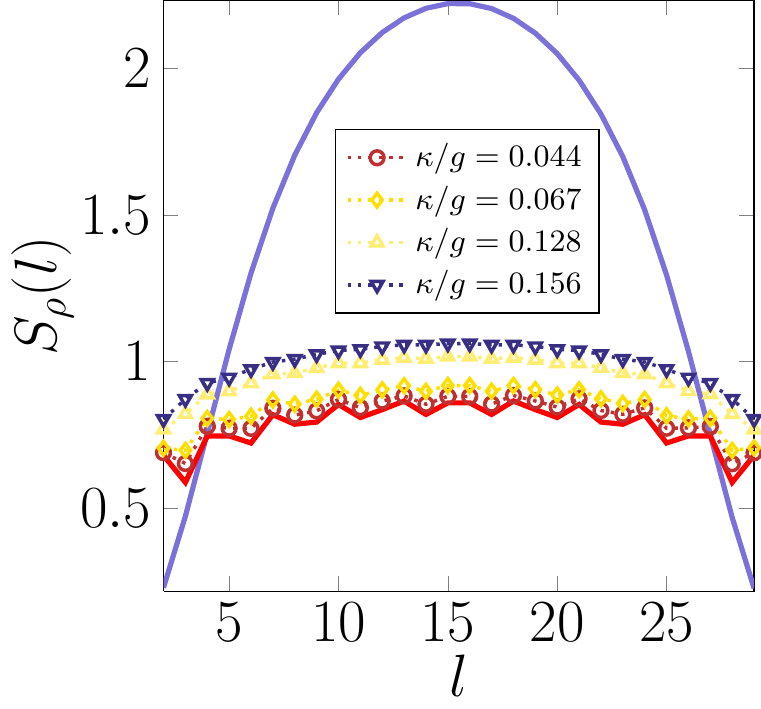}
		\label{fig:JCHentropylow}
	}
	\subfigure[][]{
		\includegraphics[height=0.208\textwidth]{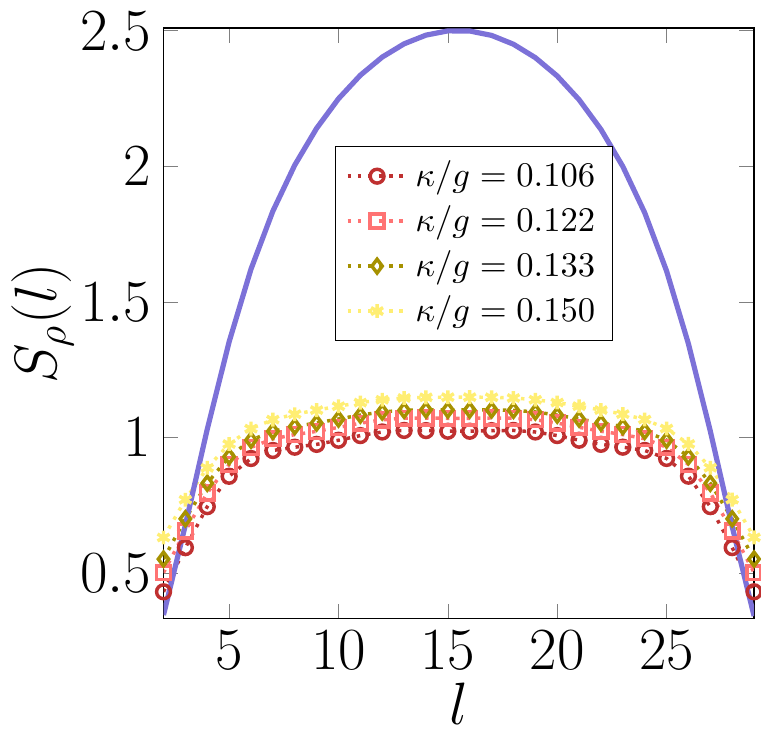}
		\label{fig:JCHentropyhigh}
	}\\
	\subfigure[][]{
		\includegraphics[height=0.200\textwidth]{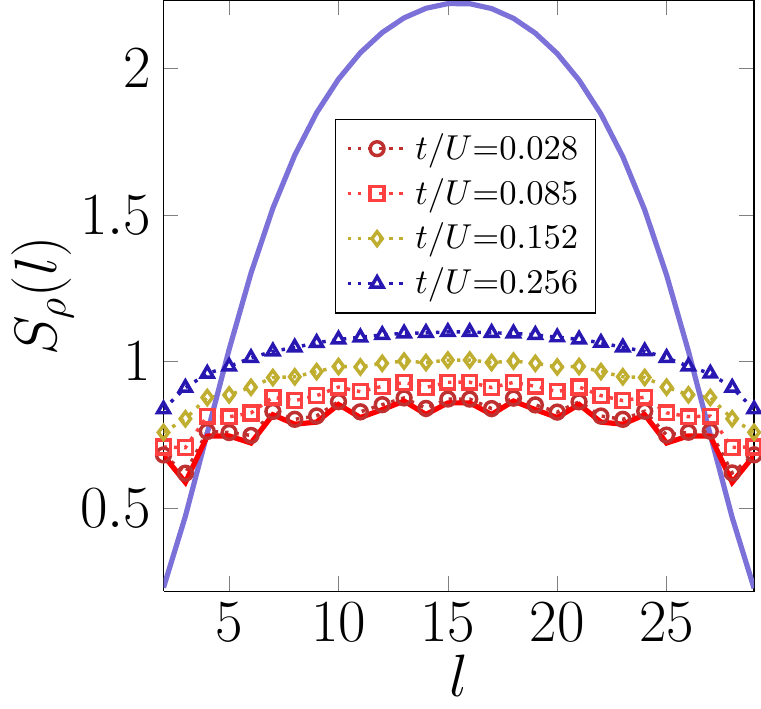}
		\label{fig:BHentropylow}
	}
	\subfigure[][]{
		\includegraphics[height=0.208\textwidth]{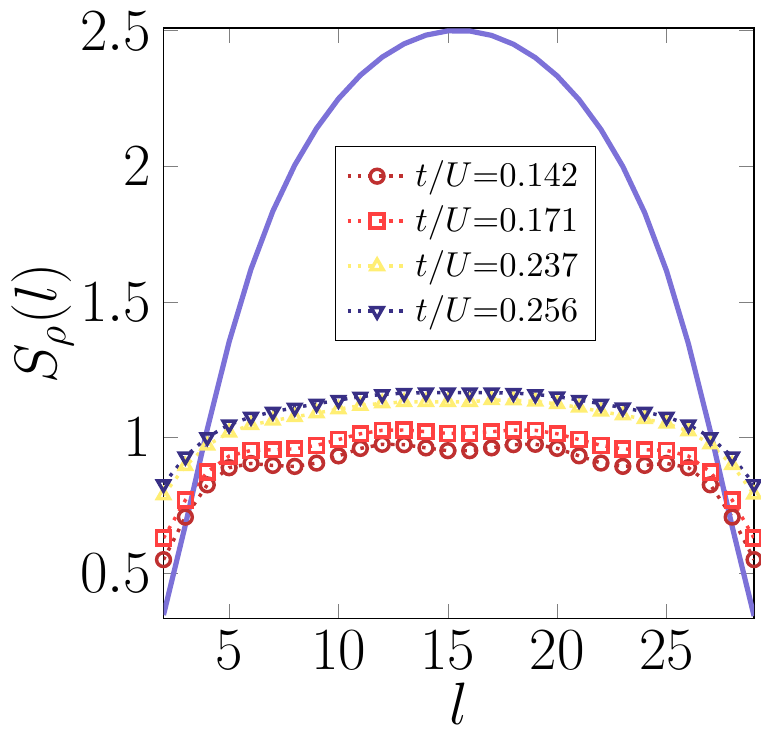}
		\label{fig:BHentropyhigh}
	}\\
\caption{The photon and boson entanglement entropy $S_{\rho}(l)$ for a 
contiguous block of sites of length $l=3$ to 29 with $L=31$, for the JCH and
BH models, respectively. Two mean densities are considered, $n=0.645$ for the 
(a) JCH and (c) BH models, and 
$n=1.129$ for the (b) JCH and (d) BH models. The solid blue line is the exact 
value of $S_{\rho}^b$ for an ideal lattice Bose gas, and the solid red line in 
(a) and (c) is the corresponding value $S_{\rho}^f$ for the ideal Fermi gas.}
	\label{fig:entropyprofile}
\end{figure}

The entanglement entropy $S_{\rho}(l)$ is plotted in 
Fig.~\ref{fig:entropyprofile} as a function of $l$ for $3\leq l\leq 29$ 
($L=31$) along a contour of constant density in the SF regime. The results are 
displayed for both the JCH and BH models at two different mean densities: 
$n=0.645$ and $n=1.129$. The corresponding entropy profiles for the free Bose 
and Fermi gases at the same mean density, calculated numerically, are plotted for 
comparison. The bosonic entanglement entropy increases approximately 
linearly with $l$ for $l\ll L$. The fermionic entanglement 
entropy displays strong oscillations but much weaker $l$-dependence. At the 
lower mean density considered, the photon entanglement entropy profile is very 
similar to the ideal Fermi gas for small hopping, again providing strong 
evidence for fermionization. As the hopping increases, the oscillations become 
less pronounced and the value of the entropy increases; presumably the entropy 
profile would converge to that of the ideal Bose gas for very large hopping
amplitudes. For the $n>1$ case, which precludes fermionization in this
single-band model, the oscillations are almost completely washed out. The
magnitude of entanglement entropy remains well below the ideal Bose gas limit,
however.

While the entanglement entropy can be used to help distinguish quantum phases,
it does not directly quantify the possible use of this entanglement for 
quantum computation. Indeed, it is difficult to conceive of how one could 
encode quantum algorithms into the (generally delocalized) indistinguishable 
bosons of the BH model. The JCH model, on the other hand, has distinguishable 
quantum registers in the spin states of the two-level atoms (i.e.\ qubits) 
which are each localized to a different optical cavity. If entanglement were
generated between cavity atoms by the itinerant photons, then the coupled
cavity QED could potentially be a natural environment for quantum computation.

It was recently proven that universal quantum computation is possible as long
as the entanglement associated with arbitrary bipartitions is 
non-vanishing~\cite{VanDenNest2013}. One useful measure that is closely 
related to the entanglement entropy is the localizable 
entanglement~\cite{Verstraete2004c,Verstraete2004b,Popp2005}. The localizable 
entanglement $E^{\text{LE}}_{ij}$ between qubits $i$ and $j$ of a multiqubit 
system is defined as the maximal entanglement that can be concentrated between 
qubits $i$ and $j$ via local (i.e.\ single-spin) operations and classical 
communication, and is a non-negative number in the range $\left[0,1\right]$. 
In a spin network, the localizable entanglement is lower-bounded by the 
maximal absolute value of the spin-spin correlation function $C^{\alpha,\beta}_{ij}
:=\langle\sigma^{\alpha}_i\sigma^{\beta}_j\rangle 
- \langle\sigma^{\alpha}_i\rangle\langle\sigma^{\beta}_j\rangle$ over all 
possible axes $\alpha,\beta\in\field{R}^3$~\cite{Popp2005,Amico2008}. Using
$\sigma^z_i=\sigma^+_i\sigma^-_i-I/2=n_i^{\rm spin}-I/2$ where $I$ is the 
$2\times 2$ identity matrix, and if $\alpha=\beta=z$, the localizable 
entanglement is at least as large as
\begin{align*}
\left | C^{zz}_{ij} \right | &\equiv \left | \langle N_i N_j\rangle 
-\langle N_i \rangle \langle N_j \rangle \right |,
\end{align*}
where here $N_k=\sigma^+_k\sigma^-_k$ is the local spin excitation number 
operator, satisfying $0\leq\langle N_k\rangle \leq 1$. The localizable 
entanglement is 
therefore closely related to the (unnormalized) two-body correlation function, 
defined in Eq.~(\ref{eq:G2}). For $i\neq j$ (the only case of interest for 
entanglement), one obtains $C^{zz}_{ij}=G^{(2)}(i,j)-\langle N_i\rangle
\langle N_j\rangle\approx G^{(2)}(i,j)-n^2$ (the local excitation number is 
approximately equal to the mean density -- the total number of excitations over
all sites -- if the density is almost constant).

The behavior of the spin-spin correlation function in the thermodynamic limit 
can be estimated under the assumption that the spin excitations have completely
fermionized, using Eq.~(\ref{eq:g2Tonks}) for a ring geometry or 
Eqs.~(\ref{eq:g2FF}-\ref{eq:g2FF2}) for open boundary conditions. Consider the 
symmetric pair of sites $i_\mp = (L\mp a)/2$, separated by $a$. Using either 
geometry one obtains $G^{(2)}(i_-,i_+) \rightarrow n^2 \sbrackets{1 
- \sin^2 (\pi n a)/(\pi n a)^2}$, assuming short-ranged correlations in the 
bulk where $a$ remains constant as $L$ increases; for small $n$, the spin-spin 
correlation function approaches $-n^2$. If the separation instead scales like 
$a \sim L$, then $G^{(2)}(i_-,i_+) \rightarrow n^2$ and $C^{zz}_{i_-i_+}
\rightarrow 0$.

The correlation function $C^{zz}_{ij}$ was calculated throughout the phase 
region investigated. Restricting $i$ and $j$ to lie between $\lceil L/4 \rceil$ 
and $\lceil 3L/4 \rceil$ to avoid boundary effects, $\left|C^{zz}_{i,j}\right|$ 
take the largest values when $|i-j|=1$ ($i=j$ is explicitly excluded). The 
values of $C^{zz}_{i,i\pm 1}$ were found to be close to zero everywhere in the 
MI phase, consistent with the strong density localization and small density 
fluctuations which are the hallmark of MI states. Likewise, the values of 
$C^{zz}_{i,i\pm 1}$ tend to zero for all $i$ at high mean densities and hopping 
where the BEC fraction is large. This reflects the smooth and relatively 
constant profile of the two-body correlation function other than the remnant 
of the exclusion hole at short distances, as shown in Fig.~\ref{fig:g2BH}. 

At low density and hopping amplitude in the SF regime, the magnitudes of 
$C^{zz}_{i,i\pm 1}$ generally range between $0.02$ to $0.05$. This is a direct 
consequence of the large exclusion hole in the spin density-density correlation 
function in the vicinity of $i\sim j$, inherited from the strongly fermionized 
polaritons. The spin-spin correlation function reaches a maximum value in the 
region directly adjacent to the hole boundary of the $n=1$ Mott lobe. At the 
point $(\tilde{\mu},\kappa/g)=(-0.989, 0.0167)$ where the mean spin excitation 
density is $n=0.3825$, the numerics yield $\left|C^{zz}_{i,i\pm 1}\right|
=0.075$. The fermionized theory predicts the comparable but slightly larger
value of $\sin^2(\pi n)/\pi^2\approx 0.088$. The difference between theory and
computation is likely partly due to the fact that the spin excitations have not
perfectly fermionized.

Because we have only considered the $zz$ quadrature of the spin-spin 
correlation function, the
numerical value is a lower-bound to the lower-bound of the localizable
entanglement. The actual value of the localizable entanglement in the 
Tonks-Girardeau regime could well be larger. In any case, the numerical 
results suggest that there is sufficient entanglement between the
atoms in the ground state of coupled cavities to support universal quantum 
computation, if a suitable strategy to embed this environment in a quantum 
circuit could be found.

\section{Conclusions}
\label{sec:conclusions}
In this work, we have explored the Mott-Insulator to superfluid transition of 
the one-dimensional Jaynes-Cummings-Hubbard (JCH) model in the strong-coupling 
regime with no detuning. The purpose of the work has been two-fold. First and
foremost, it has been to study the nature of the ground state in the vicinity 
of the phase transition, in particular to demonstrate that the photons are in 
fact strongly fermionized in the low-density `superfluid' phase. Second, it has 
been to compare and contrast the properties of the ground state to that of the 
1D Bose-Hubbard (BH) model, in order to highlight the unique features of the
JCH model. The results were obtained by finding the ground state using the 
finite-system Density Matrix Renormalization Group, computing various static 
properties, and then performing a finite-size scaling analysis to infer the 
thermodynamic limit.

The main result is that in one dimension, the ground state of the JCH model in 
the low-density regime outside the Mott-insulating lobes is dramatically 
different from that of a conventional superfluid. Rather, the system in the
region widely characterized as a superfluid (SF) is in fact a Tonks-Girardeau 
gas of strongly fermionized excitations, much as occurs in the one-dimensional 
BH model. This is evidenced by the power law of the 
single-particle density matrix in both the spin and photonic sectors of 
the model, by the Friedel oscillations and the fermionic exclusion hole in the 
two-body density matrix for each sector, and by the strongly fermionic profile
of the entanglement entropy. Thus coupled cavity QED provides a natural and 
accessible environment for the realization of strongly correlated photons. The 
photon fermionization should be readily observable in experiments using 
standard photon correlation spectroscopy~\cite{Carmichael1996}.

We have calculated the superfluid fraction for both species of excitation in 
the JCH model, both within the low-density SF regime as well as for large
tunneling beyond the expected BKT transition point, and found that the 
superfluid density vanishes in the thermodynamic limit in both cases. At the 
same time, we have found that the Bose-Einstein condensate fraction for the 
photons and spin excitations is non-vanishing throughout the SF region.
The value of the condensate fraction is low at very low densities, consistent
with fermionization, but can reach as high as 80\% at high densities for the 
photons. This indicates the existence of a (quasi)condensate of photons and
even spin excitations, despite the absence of superfluidity.

The same static properties for the 1D BH model have been calculated in an 
equivalent parameter regime in phase space, and the results have been 
compared in detail to those obtained within the JCH model.
Broadly, the behavior of the two models coincide: neither exhibits a true 
Mott-insulator to conventional superfluid transition. However, various
quantitative differences exist. The single-particle correlations in the JCH 
ground state within the $n=1$ Mott lobe decay more rapidly with increasing 
hopping that in the BH case, while those in the intermediate region between 
the $n=0$ and $n=1$ lobes decay more slowly. Consistently with this result, 
the condensate fraction for the JCH photons rises more rapidly with increasing
hopping than for the BH bosons. 
While a formal mapping between the BH and JCH models does not exist, the 
results indicate that the well-known manifestations of the BH model will be 
largely reproduced in physical systems that are well-described by the JCH
model.

That said, the JCH model has some intriguing features not shared by the BH
model, owing to the presence of two species. The spin and photon degrees of 
freedom are inextricably linked through the fundamental polariton excitations.
Thus Bose-Einstein condensation of photons implies that the spin 
excitations are similarly condensed. The atomic spin states are thereby 
effectively delocalized across the entire system in spite of the fact that each
atom is confined to its respective cavity. This observation opens the 
intriguing possibility of inducing spin liquid-like states in cavity QED 
systems. In fact, the possibility of spin dimerization in the JCH model with 
large positive detuning $\delta\gg 0$ (which induces frustration and is not
considered in the present work) has been noted very recently~\cite{Zhu2013}. 
Likewise, atoms in different cavities are spontaneously entangled via the 
itinerant photons, as evidenced by the non-zero value of the localizable spin 
entanglement in the SF regime. It would be intriguing to systematically
consider the effect of non-zero detuning on the properties of the ground 
states, but this is beyond the scope of the present work.

The majority of the parameter space explored in this project is in the vicinity 
of a phase transition, but the convergence of the DMRG algorithm is not 
guaranteed very close to the phase boundary. It could be fruitful to compare
the results with those obtained using a method such as Multiscale Entanglement 
Renormalization Ansatz~\cite{Vidal2008}, which is specifically tailored to work 
well with critical systems. In a similar vein, the calculations (in particular 
the superfluid density) could be repeated with a method that is designed for 
handling infinite systems directly, such as 
iDMRG~\cite{McCulloch2008,Crosswhite2008}.
This would avoid the need to perform finite-size scaling on small systems to 
make quantitive statements about the thermodynamic limit.

Cavity quantum electrodynamics is a promising candidate for quantum information 
processing applications, since the local atoms can be used as qubits and the 
photons can be used to generate entanglement between them. Our results suggest
that the localizable entanglement is always finite throughout the SF region.
It would be interesting to determine if the ground state for a more complex 
network of coupled cavities could be a resource for measurement-based quantum 
computation, where universal quantum algorithms are effected solely via 
single-qubit measurements conditioned on previous outcomes~\cite{Briegel2009}. 
Preliminary calculations indicate that the type of correlations between atoms 
in the current 1D JCH model with zero detuning are probably not suitable for 
gate teleportation via measurements. This will be explored more fully in future 
work.

\section{Acknowledgements}
\label{sec:acknowledgements}
The authors are grateful for research funding from the Natural Sciences and 
Engineering Research Council of Canada, Alberta Innovates -- Technology 
Futures, and the Canadian Institute for Advanced Research. This research has 
been enabled by the use of the ALPS software package, particularly the 
{\tt dmrg} application, as well as by computing resources provided by WestGrid 
and Compute/Calcul Canada.

\bibliographystyle{apsrev}
\def\urlprefix{}
\def\url#1{}
\bibliography{jch5}

\begin{thebibliography}{98}
\expandafter\ifx\csname natexlab\endcsname\relax\def\natexlab#1{#1}\fi
\expandafter\ifx\csname bibnamefont\endcsname\relax
  \def\bibnamefont#1{#1}\fi
\expandafter\ifx\csname bibfnamefont\endcsname\relax
  \def\bibfnamefont#1{#1}\fi
\expandafter\ifx\csname citenamefont\endcsname\relax
  \def\citenamefont#1{#1}\fi
\expandafter\ifx\csname url\endcsname\relax
  \def\url#1{\texttt{#1}}\fi
\expandafter\ifx\csname urlprefix\endcsname\relax\def\urlprefix{URL }\fi
\providecommand{\bibinfo}[2]{#2}
\providecommand{\eprint}[2][]{\url{#2}}

\bibitem[{\citenamefont{Fisher et~al.}(1989)\citenamefont{Fisher, Weichman,
  Grinstein, and Fisher}}]{Fisher1989}
\bibinfo{author}{\bibfnamefont{M.~P.~A.} \bibnamefont{Fisher}},
  \bibinfo{author}{\bibfnamefont{P.~B.} \bibnamefont{Weichman}},
  \bibinfo{author}{\bibfnamefont{G.}~\bibnamefont{Grinstein}},
  \bibnamefont{and} \bibinfo{author}{\bibfnamefont{D.~S.}
  \bibnamefont{Fisher}}, \bibinfo{journal}{Phys. Rev. B}
  \textbf{\bibinfo{volume}{40}}, \bibinfo{pages}{546} (\bibinfo{year}{1989}).

\bibitem[{\citenamefont{van~der Zant et~al.}(1992)\citenamefont{van~der Zant,
  Fritschy, Elion, Geerligs, and Mooij}}]{Zant-1992}
\bibinfo{author}{\bibfnamefont{H.~S.~J.} \bibnamefont{van~der Zant}},
  \bibinfo{author}{\bibfnamefont{F.~C.} \bibnamefont{Fritschy}},
  \bibinfo{author}{\bibfnamefont{W.~J.} \bibnamefont{Elion}},
  \bibinfo{author}{\bibfnamefont{L.~J.} \bibnamefont{Geerligs}},
  \bibnamefont{and} \bibinfo{author}{\bibfnamefont{J.~E.} \bibnamefont{Mooij}},
  \bibinfo{journal}{Phys. Rev. Lett.} \textbf{\bibinfo{volume}{69}},
  \bibinfo{pages}{2971} (\bibinfo{year}{1992}).

\bibitem[{\citenamefont{Bloch}(2005)}]{Bloch2005}
\bibinfo{author}{\bibfnamefont{I.}~\bibnamefont{Bloch}}, \bibinfo{journal}{Nat.
  Phys.} \textbf{\bibinfo{volume}{1}}, \bibinfo{pages}{23}
  (\bibinfo{year}{2005}).

\bibitem[{\citenamefont{Corrielli et~al.}(2013)\citenamefont{Corrielli, Crespi,
  Valle, Longhi, and Osellame}}]{Corrielli2013}
\bibinfo{author}{\bibfnamefont{G.}~\bibnamefont{Corrielli}},
  \bibinfo{author}{\bibfnamefont{A.}~\bibnamefont{Crespi}},
  \bibinfo{author}{\bibfnamefont{G.~D.} \bibnamefont{Valle}},
  \bibinfo{author}{\bibfnamefont{S.}~\bibnamefont{Longhi}}, \bibnamefont{and}
  \bibinfo{author}{\bibfnamefont{R.}~\bibnamefont{Osellame}},
  \bibinfo{journal}{Nat. Commun.} \textbf{\bibinfo{volume}{4}},
  \bibinfo{pages}{1555} (\bibinfo{year}{2013}).

\bibitem[{\citenamefont{Hartmann et~al.}(2006)\citenamefont{Hartmann,
  Brand\~{a}o, and Plenio}}]{Hartmann2006}
\bibinfo{author}{\bibfnamefont{M.~J.} \bibnamefont{Hartmann}},
  \bibinfo{author}{\bibfnamefont{F.~G. S.~L.} \bibnamefont{Brand\~{a}o}},
  \bibnamefont{and} \bibinfo{author}{\bibfnamefont{M.~B.}
  \bibnamefont{Plenio}}, \bibinfo{journal}{Nat. Phys.}
  \textbf{\bibinfo{volume}{2}}, \bibinfo{pages}{849 } (\bibinfo{year}{2006}).

\bibitem[{\citenamefont{Greentree et~al.}(2006)\citenamefont{Greentree, Tahan,
  Cole, and Hollenberg}}]{Greentree2006}
\bibinfo{author}{\bibfnamefont{A.~D.} \bibnamefont{Greentree}},
  \bibinfo{author}{\bibfnamefont{C.}~\bibnamefont{Tahan}},
  \bibinfo{author}{\bibfnamefont{J.~H.} \bibnamefont{Cole}}, \bibnamefont{and}
  \bibinfo{author}{\bibfnamefont{L.~C.~L.} \bibnamefont{Hollenberg}},
  \bibinfo{journal}{Nat. Phys.} \textbf{\bibinfo{volume}{2}},
  \bibinfo{pages}{856 } (\bibinfo{year}{2006}).

\bibitem[{\citenamefont{Angelakis et~al.}(2007)\citenamefont{Angelakis, Santos,
  and Bose}}]{Angelakis2007}
\bibinfo{author}{\bibfnamefont{D.~G.} \bibnamefont{Angelakis}},
  \bibinfo{author}{\bibfnamefont{M.~F.} \bibnamefont{Santos}},
  \bibnamefont{and} \bibinfo{author}{\bibfnamefont{S.}~\bibnamefont{Bose}},
  \bibinfo{journal}{Phys. Rev. A} \textbf{\bibinfo{volume}{76}},
  \bibinfo{pages}{031805} (\bibinfo{year}{2007}).

\bibitem[{\citenamefont{Houck et~al.}(2012)\citenamefont{Houck, T\"{u}reci, and
  Koch}}]{Houck2012}
\bibinfo{author}{\bibfnamefont{A.~A.} \bibnamefont{Houck}},
  \bibinfo{author}{\bibfnamefont{H.~E.} \bibnamefont{T\"{u}reci}},
  \bibnamefont{and} \bibinfo{author}{\bibfnamefont{J.}~\bibnamefont{Koch}},
  \bibinfo{journal}{Nat. Phys.} \textbf{\bibinfo{volume}{8}},
  \bibinfo{pages}{292} (\bibinfo{year}{2012}).

\bibitem[{\citenamefont{Mott and Peierls}(1937)}]{Mott1937}
\bibinfo{author}{\bibfnamefont{N.~F.} \bibnamefont{Mott}} \bibnamefont{and}
  \bibinfo{author}{\bibfnamefont{R.}~\bibnamefont{Peierls}},
  \bibinfo{journal}{Proc. Phys. Soc.} \textbf{\bibinfo{volume}{49}},
  \bibinfo{pages}{72} (\bibinfo{year}{1937}).

\bibitem[{\citenamefont{Mott}(1949)}]{Mott1949}
\bibinfo{author}{\bibfnamefont{N.~F.} \bibnamefont{Mott}},
  \bibinfo{journal}{Proc. Phys. Soc. A} \textbf{\bibinfo{volume}{62}},
  \bibinfo{pages}{416} (\bibinfo{year}{1949}).

\bibitem[{\citenamefont{Mott}(1982)}]{Mott1982}
\bibinfo{author}{\bibfnamefont{N.~F.} \bibnamefont{Mott}},
  \bibinfo{journal}{Proc. Roy. Soc. London. A.} \textbf{\bibinfo{volume}{382}},
  \bibinfo{pages}{1} (\bibinfo{year}{1982}).

\bibitem[{\citenamefont{Kapitza}(1938)}]{Kapitza1938}
\bibinfo{author}{\bibfnamefont{P.}~\bibnamefont{Kapitza}},
  \bibinfo{journal}{Nature} \textbf{\bibinfo{volume}{141}}, \bibinfo{pages}{74}
  (\bibinfo{year}{1938}).

\bibitem[{\citenamefont{Allen and Misener}(1938)}]{Allen1938}
\bibinfo{author}{\bibfnamefont{J.~F.} \bibnamefont{Allen}} \bibnamefont{and}
  \bibinfo{author}{\bibfnamefont{A.~D.} \bibnamefont{Misener}},
  \bibinfo{journal}{Nature} \textbf{\bibinfo{volume}{141}}, \bibinfo{pages}{75}
  (\bibinfo{year}{1938}).

\bibitem[{\citenamefont{Landau}(1941)}]{Landau1941}
\bibinfo{author}{\bibfnamefont{L.~D.} \bibnamefont{Landau}},
  \bibinfo{journal}{J. Phys. USSR} \textbf{\bibinfo{volume}{5}},
  \bibinfo{pages}{71} (\bibinfo{year}{1941}).

\bibitem[{\citenamefont{Greiner et~al.}(2002)\citenamefont{Greiner, Mandel,
  Esslinger, H\"{a}nsch, and Bloch}}]{Greiner2002}
\bibinfo{author}{\bibfnamefont{M.}~\bibnamefont{Greiner}},
  \bibinfo{author}{\bibfnamefont{O.}~\bibnamefont{Mandel}},
  \bibinfo{author}{\bibfnamefont{T.}~\bibnamefont{Esslinger}},
  \bibinfo{author}{\bibfnamefont{T.~W.} \bibnamefont{H\"{a}nsch}},
  \bibnamefont{and} \bibinfo{author}{\bibfnamefont{I.}~\bibnamefont{Bloch}},
  \bibinfo{journal}{Nature} \textbf{\bibinfo{volume}{415}}, \bibinfo{pages}{39}
  (\bibinfo{year}{2002}).

\bibitem[{\citenamefont{St\"oferle et~al.}(2004)\citenamefont{St\"oferle,
  Moritz, Schori, K\"ohl, and Esslinger}}]{Stoferle2004}
\bibinfo{author}{\bibfnamefont{T.}~\bibnamefont{St\"oferle}},
  \bibinfo{author}{\bibfnamefont{H.}~\bibnamefont{Moritz}},
  \bibinfo{author}{\bibfnamefont{C.}~\bibnamefont{Schori}},
  \bibinfo{author}{\bibfnamefont{M.}~\bibnamefont{K\"ohl}}, \bibnamefont{and}
  \bibinfo{author}{\bibfnamefont{T.}~\bibnamefont{Esslinger}},
  \bibinfo{journal}{Phys. Rev. Lett.} \textbf{\bibinfo{volume}{92}},
  \bibinfo{pages}{130403} (\bibinfo{year}{2004}).

\bibitem[{\citenamefont{Spielman et~al.}(2007)\citenamefont{Spielman, Phillips,
  and Porto}}]{Spielman2007}
\bibinfo{author}{\bibfnamefont{I.~B.} \bibnamefont{Spielman}},
  \bibinfo{author}{\bibfnamefont{W.~D.} \bibnamefont{Phillips}},
  \bibnamefont{and} \bibinfo{author}{\bibfnamefont{J.~V.} \bibnamefont{Porto}},
  \bibinfo{journal}{Phys. Rev. Lett.} \textbf{\bibinfo{volume}{98}},
  \bibinfo{pages}{080404} (\bibinfo{year}{2007}).

\bibitem[{\citenamefont{Gemelke et~al.}(2009)\citenamefont{Gemelke, Zhang,
  Hung, and Chin}}]{Gemelke2009}
\bibinfo{author}{\bibfnamefont{N.}~\bibnamefont{Gemelke}},
  \bibinfo{author}{\bibfnamefont{X.}~\bibnamefont{Zhang}},
  \bibinfo{author}{\bibfnamefont{C.-L.} \bibnamefont{Hung}}, \bibnamefont{and}
  \bibinfo{author}{\bibfnamefont{C.}~\bibnamefont{Chin}},
  \bibinfo{journal}{Nature} \textbf{\bibinfo{volume}{460}},
  \bibinfo{pages}{995} (\bibinfo{year}{2009}).

\bibitem[{\citenamefont{Bakr et~al.}(2010)\citenamefont{Bakr, Peng, Tai, Ma,
  Simon, Gillen, Foelling, Pollet, and Greiner}}]{Bakr2010}
\bibinfo{author}{\bibfnamefont{W.~S.} \bibnamefont{Bakr}},
  \bibinfo{author}{\bibfnamefont{A.}~\bibnamefont{Peng}},
  \bibinfo{author}{\bibfnamefont{M.~E.} \bibnamefont{Tai}},
  \bibinfo{author}{\bibfnamefont{R.}~\bibnamefont{Ma}},
  \bibinfo{author}{\bibfnamefont{J.}~\bibnamefont{Simon}},
  \bibinfo{author}{\bibfnamefont{J.~I.} \bibnamefont{Gillen}},
  \bibinfo{author}{\bibfnamefont{S.}~\bibnamefont{Foelling}},
  \bibinfo{author}{\bibfnamefont{L.}~\bibnamefont{Pollet}}, \bibnamefont{and}
  \bibinfo{author}{\bibfnamefont{M.}~\bibnamefont{Greiner}},
  \bibinfo{journal}{Science} \textbf{\bibinfo{volume}{329}},
  \bibinfo{pages}{547} (\bibinfo{year}{2010}).

\bibitem[{\citenamefont{Haller et~al.}(2010)\citenamefont{Haller, Hart, Mark,
  Danzl, Reichsoellner, Gustavsson, Dalmonte, Pupillo, and
  Naegerl}}]{Haller2010}
\bibinfo{author}{\bibfnamefont{E.}~\bibnamefont{Haller}},
  \bibinfo{author}{\bibfnamefont{R.}~\bibnamefont{Hart}},
  \bibinfo{author}{\bibfnamefont{M.~J.} \bibnamefont{Mark}},
  \bibinfo{author}{\bibfnamefont{J.~G.} \bibnamefont{Danzl}},
  \bibinfo{author}{\bibfnamefont{L.}~\bibnamefont{Reichsoellner}},
  \bibinfo{author}{\bibfnamefont{M.}~\bibnamefont{Gustavsson}},
  \bibinfo{author}{\bibfnamefont{M.}~\bibnamefont{Dalmonte}},
  \bibinfo{author}{\bibfnamefont{G.}~\bibnamefont{Pupillo}}, \bibnamefont{and}
  \bibinfo{author}{\bibfnamefont{H.-C.} \bibnamefont{Naegerl}},
  \bibinfo{journal}{Nature} \textbf{\bibinfo{volume}{466}},
  \bibinfo{pages}{597} (\bibinfo{year}{2010}).

\bibitem[{\citenamefont{Trotzky et~al.}(2010)\citenamefont{Trotzky, Pollet,
  Gerbier, Schnorrberger, Bloch, Prokof'ev, Svistunov, and
  Troyer}}]{Trotzky2010}
\bibinfo{author}{\bibfnamefont{S.}~\bibnamefont{Trotzky}},
  \bibinfo{author}{\bibfnamefont{L.}~\bibnamefont{Pollet}},
  \bibinfo{author}{\bibfnamefont{F.}~\bibnamefont{Gerbier}},
  \bibinfo{author}{\bibfnamefont{U.}~\bibnamefont{Schnorrberger}},
  \bibinfo{author}{\bibfnamefont{I.}~\bibnamefont{Bloch}},
  \bibinfo{author}{\bibfnamefont{N.~V.} \bibnamefont{Prokof'ev}},
  \bibinfo{author}{\bibfnamefont{B.}~\bibnamefont{Svistunov}},
  \bibnamefont{and} \bibinfo{author}{\bibfnamefont{M.}~\bibnamefont{Troyer}},
  \bibinfo{journal}{Nat. Phys.} \textbf{\bibinfo{volume}{6}},
  \bibinfo{pages}{998} (\bibinfo{year}{2010}).

\bibitem[{\citenamefont{Hartmann et~al.}(2007)\citenamefont{Hartmann, Brandao,
  and Plenio}}]{Hartmann2007}
\bibinfo{author}{\bibfnamefont{M.~J.} \bibnamefont{Hartmann}},
  \bibinfo{author}{\bibfnamefont{F.~G. S.~L.} \bibnamefont{Brandao}},
  \bibnamefont{and} \bibinfo{author}{\bibfnamefont{M.~B.}
  \bibnamefont{Plenio}}, \bibinfo{journal}{Phys. Rev. Lett.}
  \textbf{\bibinfo{volume}{99}}, \bibinfo{pages}{160501}
  (\bibinfo{year}{2007}).

\bibitem[{\citenamefont{Rossini and Fazio}(2007)}]{Rossini2007}
\bibinfo{author}{\bibfnamefont{D.}~\bibnamefont{Rossini}} \bibnamefont{and}
  \bibinfo{author}{\bibfnamefont{R.}~\bibnamefont{Fazio}},
  \bibinfo{journal}{Phys. Rev. Lett.} \textbf{\bibinfo{volume}{99}},
  \bibinfo{pages}{186401} (\bibinfo{year}{2007}).

\bibitem[{\citenamefont{Carusotto et~al.}(2009)\citenamefont{Carusotto, Gerace,
  Tureci, De~Liberato, Ciuti, and Imamo\ifmmode~\check{g}\else
  \v{g}\fi{}lu}}]{Carusotto2009}
\bibinfo{author}{\bibfnamefont{I.}~\bibnamefont{Carusotto}},
  \bibinfo{author}{\bibfnamefont{D.}~\bibnamefont{Gerace}},
  \bibinfo{author}{\bibfnamefont{H.~E.} \bibnamefont{Tureci}},
  \bibinfo{author}{\bibfnamefont{S.}~\bibnamefont{De~Liberato}},
  \bibinfo{author}{\bibfnamefont{C.}~\bibnamefont{Ciuti}}, \bibnamefont{and}
  \bibinfo{author}{\bibfnamefont{A.}~\bibnamefont{Imamo\ifmmode~\check{g}\else
  \v{g}\fi{}lu}}, \bibinfo{journal}{Phys. Rev. Lett.}
  \textbf{\bibinfo{volume}{103}}, \bibinfo{pages}{033601}
  (\bibinfo{year}{2009}).

\bibitem[{\citenamefont{Kiffner and Hartmann}(2010)}]{Kiffner2010}
\bibinfo{author}{\bibfnamefont{M.}~\bibnamefont{Kiffner}} \bibnamefont{and}
  \bibinfo{author}{\bibfnamefont{M.~J.} \bibnamefont{Hartmann}},
  \bibinfo{journal}{Phys. Rev. A} \textbf{\bibinfo{volume}{81}},
  \bibinfo{pages}{021806} (\bibinfo{year}{2010}).

\bibitem[{\citenamefont{Halu et~al.}(2013)\citenamefont{Halu, Garnerone,
  Vezzani, and Bianconi}}]{Halu2013}
\bibinfo{author}{\bibfnamefont{A.}~\bibnamefont{Halu}},
  \bibinfo{author}{\bibfnamefont{S.}~\bibnamefont{Garnerone}},
  \bibinfo{author}{\bibfnamefont{A.}~\bibnamefont{Vezzani}}, \bibnamefont{and}
  \bibinfo{author}{\bibfnamefont{G.}~\bibnamefont{Bianconi}},
  \bibinfo{journal}{Phys. Rev. E} \textbf{\bibinfo{volume}{87}},
  \bibinfo{pages}{022104} (\bibinfo{year}{2013}).

\bibitem[{\citenamefont{Hayward et~al.}(2012)\citenamefont{Hayward, Martin, and
  Greentree}}]{Hayward2012}
\bibinfo{author}{\bibfnamefont{A.~L.~C.} \bibnamefont{Hayward}},
  \bibinfo{author}{\bibfnamefont{A.~M.} \bibnamefont{Martin}},
  \bibnamefont{and} \bibinfo{author}{\bibfnamefont{A.~D.}
  \bibnamefont{Greentree}}, \bibinfo{journal}{Phys. Rev. Lett.}
  \textbf{\bibinfo{volume}{108}}, \bibinfo{pages}{223602}
  (\bibinfo{year}{2012}).

\bibitem[{\citenamefont{Schir\'o et~al.}(2012)\citenamefont{Schir\'o, Bordyuh,
  \"Oztop, and T\"ureci}}]{Schiro2012}
\bibinfo{author}{\bibfnamefont{M.}~\bibnamefont{Schir\'o}},
  \bibinfo{author}{\bibfnamefont{M.}~\bibnamefont{Bordyuh}},
  \bibinfo{author}{\bibfnamefont{B.}~\bibnamefont{\"Oztop}}, \bibnamefont{and}
  \bibinfo{author}{\bibfnamefont{H.~E.} \bibnamefont{T\"ureci}},
  \bibinfo{journal}{Phys. Rev. Lett.} \textbf{\bibinfo{volume}{109}},
  \bibinfo{pages}{053601} (\bibinfo{year}{2012}).

\bibitem[{\citenamefont{Hartmann et~al.}(2008)\citenamefont{Hartmann,
  Brand{\~a}o, and Plenio}}]{Hartmann2008a}
\bibinfo{author}{\bibfnamefont{M.~J.} \bibnamefont{Hartmann}},
  \bibinfo{author}{\bibfnamefont{F.~G. S.~L.} \bibnamefont{Brand{\~a}o}},
  \bibnamefont{and} \bibinfo{author}{\bibfnamefont{M.~B.}
  \bibnamefont{Plenio}}, \bibinfo{journal}{New J. Phys.}
  \textbf{\bibinfo{volume}{10}}, \bibinfo{pages}{033011}
  (\bibinfo{year}{2008}).

\bibitem[{\citenamefont{Makin et~al.}(2008)\citenamefont{Makin, Cole, Tahan,
  Hollenberg, and Greentree}}]{Makin2008}
\bibinfo{author}{\bibfnamefont{M.~I.} \bibnamefont{Makin}},
  \bibinfo{author}{\bibfnamefont{J.~H.} \bibnamefont{Cole}},
  \bibinfo{author}{\bibfnamefont{C.}~\bibnamefont{Tahan}},
  \bibinfo{author}{\bibfnamefont{L.~C.~L.} \bibnamefont{Hollenberg}},
  \bibnamefont{and} \bibinfo{author}{\bibfnamefont{A.~D.}
  \bibnamefont{Greentree}}, \bibinfo{journal}{Phys. Rev. A}
  \textbf{\bibinfo{volume}{77}}, \bibinfo{pages}{053819}
  (\bibinfo{year}{2008}).

\bibitem[{\citenamefont{Jaynes and Cummings}(1963)}]{Jaynes1963}
\bibinfo{author}{\bibfnamefont{E.}~\bibnamefont{Jaynes}} \bibnamefont{and}
  \bibinfo{author}{\bibfnamefont{F.}~\bibnamefont{Cummings}},
  \bibinfo{journal}{Proc. IEEE} \textbf{\bibinfo{volume}{51}},
  \bibinfo{pages}{89 } (\bibinfo{year}{1963}).

\bibitem[{\citenamefont{Birnbaum et~al.}(2005)\citenamefont{Birnbaum, Boca,
  Miller, Boozer, Northup, and Kimble}}]{Birnbaum2005}
\bibinfo{author}{\bibfnamefont{K.}~\bibnamefont{Birnbaum}},
  \bibinfo{author}{\bibfnamefont{A.}~\bibnamefont{Boca}},
  \bibinfo{author}{\bibfnamefont{R.}~\bibnamefont{Miller}},
  \bibinfo{author}{\bibfnamefont{A.}~\bibnamefont{Boozer}},
  \bibinfo{author}{\bibfnamefont{T.}~\bibnamefont{Northup}}, \bibnamefont{and}
  \bibinfo{author}{\bibfnamefont{H.}~\bibnamefont{Kimble}},
  \bibinfo{journal}{Nature} \textbf{\bibinfo{volume}{436}}, \bibinfo{pages}{87}
  (\bibinfo{year}{2005}).

\bibitem[{\citenamefont{Aichhorn et~al.}(2008)\citenamefont{Aichhorn,
  Hohenadler, Tahan, and Littlewood}}]{Aichhorn2008}
\bibinfo{author}{\bibfnamefont{M.}~\bibnamefont{Aichhorn}},
  \bibinfo{author}{\bibfnamefont{M.}~\bibnamefont{Hohenadler}},
  \bibinfo{author}{\bibfnamefont{C.}~\bibnamefont{Tahan}}, \bibnamefont{and}
  \bibinfo{author}{\bibfnamefont{P.~B.} \bibnamefont{Littlewood}},
  \bibinfo{journal}{Phys. Rev. Lett.} \textbf{\bibinfo{volume}{100}},
  \bibinfo{pages}{216401} (\bibinfo{year}{2008}).

\bibitem[{\citenamefont{Koch and LeHur}(2009)}]{Koch2009}
\bibinfo{author}{\bibfnamefont{J.}~\bibnamefont{Koch}} \bibnamefont{and}
  \bibinfo{author}{\bibfnamefont{K.}~\bibnamefont{LeHur}},
  \bibinfo{journal}{Phys. Rev. A} \textbf{\bibinfo{volume}{80}},
  \bibinfo{pages}{023811} (\bibinfo{year}{2009}).

\bibitem[{\citenamefont{Schmidt and Blatter}(2009)}]{Schmidt2009}
\bibinfo{author}{\bibfnamefont{S.}~\bibnamefont{Schmidt}} \bibnamefont{and}
  \bibinfo{author}{\bibfnamefont{G.}~\bibnamefont{Blatter}},
  \bibinfo{journal}{Phys. Rev. Lett.} \textbf{\bibinfo{volume}{103}},
  \bibinfo{pages}{086403} (\bibinfo{year}{2009}).

\bibitem[{\citenamefont{Pippan et~al.}(2009)\citenamefont{Pippan, Evertz, and
  Hohenadler}}]{Pippan2009}
\bibinfo{author}{\bibfnamefont{P.}~\bibnamefont{Pippan}},
  \bibinfo{author}{\bibfnamefont{H.~G.} \bibnamefont{Evertz}},
  \bibnamefont{and}
  \bibinfo{author}{\bibfnamefont{M.}~\bibnamefont{Hohenadler}},
  \bibinfo{journal}{Phys. Rev. A} \textbf{\bibinfo{volume}{80}},
  \bibinfo{pages}{033612} (\bibinfo{year}{2009}).

\bibitem[{\citenamefont{Schmidt and Blatter}(2010)}]{Schmidt2010}
\bibinfo{author}{\bibfnamefont{S.}~\bibnamefont{Schmidt}} \bibnamefont{and}
  \bibinfo{author}{\bibfnamefont{G.}~\bibnamefont{Blatter}},
  \bibinfo{journal}{Phys. Rev. Lett.} \textbf{\bibinfo{volume}{104}},
  \bibinfo{pages}{216402} (\bibinfo{year}{2010}).

\bibitem[{\citenamefont{Knap et~al.}(2010)\citenamefont{Knap, Arrigoni, and
  von~der Linden}}]{Knap2010}
\bibinfo{author}{\bibfnamefont{M.}~\bibnamefont{Knap}},
  \bibinfo{author}{\bibfnamefont{E.}~\bibnamefont{Arrigoni}}, \bibnamefont{and}
  \bibinfo{author}{\bibfnamefont{W.}~\bibnamefont{von~der Linden}},
  \bibinfo{journal}{Phys. Rev. B} \textbf{\bibinfo{volume}{81}},
  \bibinfo{pages}{104303} (\bibinfo{year}{2010}).

\bibitem[{\citenamefont{Hohenadler et~al.}(2011)\citenamefont{Hohenadler,
  Aichhorn, Schmidt, and Pollet}}]{Hohenadler2011}
\bibinfo{author}{\bibfnamefont{M.}~\bibnamefont{Hohenadler}},
  \bibinfo{author}{\bibfnamefont{M.}~\bibnamefont{Aichhorn}},
  \bibinfo{author}{\bibfnamefont{S.}~\bibnamefont{Schmidt}}, \bibnamefont{and}
  \bibinfo{author}{\bibfnamefont{L.}~\bibnamefont{Pollet}},
  \bibinfo{journal}{Phys. Rev. A} \textbf{\bibinfo{volume}{84}},
  \bibinfo{pages}{041608(R)} (\bibinfo{year}{2011}).

\bibitem[{\citenamefont{Na et~al.}(2008)\citenamefont{Na, Utsunomiya, Tian, and
  Yamamoto}}]{Na2008}
\bibinfo{author}{\bibfnamefont{N.}~\bibnamefont{Na}},
  \bibinfo{author}{\bibfnamefont{S.}~\bibnamefont{Utsunomiya}},
  \bibinfo{author}{\bibfnamefont{L.}~\bibnamefont{Tian}}, \bibnamefont{and}
  \bibinfo{author}{\bibfnamefont{Y.}~\bibnamefont{Yamamoto}},
  \bibinfo{journal}{Phys. Rev. A} \textbf{\bibinfo{volume}{77}},
  \bibinfo{pages}{031803(R)} (\bibinfo{year}{2008}).

\bibitem[{\citenamefont{Ivanov et~al.}(2009)\citenamefont{Ivanov, Ivanov,
  Vitanov, Mering, Fleischhauer, and Singer}}]{Ivanov2009}
\bibinfo{author}{\bibfnamefont{P.~A.} \bibnamefont{Ivanov}},
  \bibinfo{author}{\bibfnamefont{S.~S.} \bibnamefont{Ivanov}},
  \bibinfo{author}{\bibfnamefont{N.~V.} \bibnamefont{Vitanov}},
  \bibinfo{author}{\bibfnamefont{A.}~\bibnamefont{Mering}},
  \bibinfo{author}{\bibfnamefont{M.}~\bibnamefont{Fleischhauer}},
  \bibnamefont{and} \bibinfo{author}{\bibfnamefont{K.}~\bibnamefont{Singer}},
  \bibinfo{journal}{Phys. Rev. A} \textbf{\bibinfo{volume}{80}},
  \bibinfo{pages}{060301(R)} (\bibinfo{year}{2009}).

\bibitem[{\citenamefont{Mering et~al.}(2009)\citenamefont{Mering, Fleischhauer,
  Ivanov, and Singer}}]{Mering2009}
\bibinfo{author}{\bibfnamefont{A.}~\bibnamefont{Mering}},
  \bibinfo{author}{\bibfnamefont{M.}~\bibnamefont{Fleischhauer}},
  \bibinfo{author}{\bibfnamefont{P.~A.} \bibnamefont{Ivanov}},
  \bibnamefont{and} \bibinfo{author}{\bibfnamefont{K.}~\bibnamefont{Singer}},
  \bibinfo{journal}{Phys. Rev. A} \textbf{\bibinfo{volume}{80}},
  \bibinfo{pages}{053821} (\bibinfo{year}{2009}).

\bibitem[{\citenamefont{Nunnenkamp et~al.}(2011)\citenamefont{Nunnenkamp, Koch,
  and Girvin}}]{Nunnenkamp2011}
\bibinfo{author}{\bibfnamefont{A.}~\bibnamefont{Nunnenkamp}},
  \bibinfo{author}{\bibfnamefont{J.}~\bibnamefont{Koch}}, \bibnamefont{and}
  \bibinfo{author}{\bibfnamefont{S.~M.} \bibnamefont{Girvin}},
  \bibinfo{journal}{New J. Phys.} \textbf{\bibinfo{volume}{13}},
  \bibinfo{pages}{095008} (\bibinfo{year}{2011}).

\bibitem[{\citenamefont{Wu et~al.}(2011)\citenamefont{Wu, Gao, Deng, Dai, Chen,
  and Li}}]{Wu2011}
\bibinfo{author}{\bibfnamefont{C.-W.} \bibnamefont{Wu}},
  \bibinfo{author}{\bibfnamefont{M.}~\bibnamefont{Gao}},
  \bibinfo{author}{\bibfnamefont{Z.-J.} \bibnamefont{Deng}},
  \bibinfo{author}{\bibfnamefont{H.-Y.} \bibnamefont{Dai}},
  \bibinfo{author}{\bibfnamefont{P.-X.} \bibnamefont{Chen}}, \bibnamefont{and}
  \bibinfo{author}{\bibfnamefont{C.-Z.} \bibnamefont{Li}},
  \bibinfo{journal}{Phys. Rev. A} \textbf{\bibinfo{volume}{84}},
  \bibinfo{pages}{043827} (\bibinfo{year}{2011}).

\bibitem[{\citenamefont{Schmidt and Koch}(2013)}]{Schmidt2013}
\bibinfo{author}{\bibfnamefont{S.}~\bibnamefont{Schmidt}} \bibnamefont{and}
  \bibinfo{author}{\bibfnamefont{J.}~\bibnamefont{Koch}},
  \bibinfo{journal}{Ann. Phys.} \textbf{\bibinfo{volume}{525}},
  \bibinfo{pages}{395} (\bibinfo{year}{2013}).

\bibitem[{\citenamefont{Tonks}(1936)}]{Tonks1936}
\bibinfo{author}{\bibfnamefont{L.}~\bibnamefont{Tonks}},
  \bibinfo{journal}{Phys. Rev.} \textbf{\bibinfo{volume}{50}},
  \bibinfo{pages}{955} (\bibinfo{year}{1936}).

\bibitem[{\citenamefont{Girardeau}(1960)}]{Girardeau1960}
\bibinfo{author}{\bibfnamefont{M.}~\bibnamefont{Girardeau}},
  \bibinfo{journal}{J. Math. Phys.} \textbf{\bibinfo{volume}{1}},
  \bibinfo{pages}{516} (\bibinfo{year}{1960}).

\bibitem[{\citenamefont{Lieb and Liniger}(1963)}]{Lieb1963}
\bibinfo{author}{\bibfnamefont{E.~H.} \bibnamefont{Lieb}} \bibnamefont{and}
  \bibinfo{author}{\bibfnamefont{W.}~\bibnamefont{Liniger}},
  \bibinfo{journal}{Phys. Rev.} \textbf{\bibinfo{volume}{130}},
  \bibinfo{pages}{1605} (\bibinfo{year}{1963}).

\bibitem[{\citenamefont{Giamarchi}(2003)}]{Giamarchi2003}
\bibinfo{author}{\bibfnamefont{T.}~\bibnamefont{Giamarchi}},
  \emph{\bibinfo{title}{Quantum Physics in One Dimension}}
  (\bibinfo{publisher}{Oxford University Press}, \bibinfo{year}{2003}).

\bibitem[{\citenamefont{Paredes et~al.}(2004)\citenamefont{Paredes, Widera,
  Murg, Mandel, F\"{o}lling, Cirac, Shlyapnikov, H\"{a}nsch, and
  Bloch}}]{Paredes2004}
\bibinfo{author}{\bibfnamefont{B.}~\bibnamefont{Paredes}},
  \bibinfo{author}{\bibfnamefont{A.}~\bibnamefont{Widera}},
  \bibinfo{author}{\bibfnamefont{V.}~\bibnamefont{Murg}},
  \bibinfo{author}{\bibfnamefont{O.}~\bibnamefont{Mandel}},
  \bibinfo{author}{\bibfnamefont{S.}~\bibnamefont{F\"{o}lling}},
  \bibinfo{author}{\bibfnamefont{I.}~\bibnamefont{Cirac}},
  \bibinfo{author}{\bibfnamefont{G.~V.} \bibnamefont{Shlyapnikov}},
  \bibinfo{author}{\bibfnamefont{T.~W.} \bibnamefont{H\"{a}nsch}},
  \bibnamefont{and} \bibinfo{author}{\bibfnamefont{I.}~\bibnamefont{Bloch}},
  \bibinfo{journal}{Nature} \textbf{\bibinfo{volume}{429}},
  \bibinfo{pages}{277} (\bibinfo{year}{2004}).

\bibitem[{\citenamefont{Kinoshita et~al.}(2004)\citenamefont{Kinoshita, Wenger,
  and Weiss}}]{Kinoshita2004}
\bibinfo{author}{\bibfnamefont{T.}~\bibnamefont{Kinoshita}},
  \bibinfo{author}{\bibfnamefont{T.}~\bibnamefont{Wenger}}, \bibnamefont{and}
  \bibinfo{author}{\bibfnamefont{D.~S.} \bibnamefont{Weiss}},
  \bibinfo{journal}{Science} \textbf{\bibinfo{volume}{305}},
  \bibinfo{pages}{1125} (\bibinfo{year}{2004}).

\bibitem[{\citenamefont{Mandel and Wolf}(1995)}]{Mandel1995}
\bibinfo{author}{\bibfnamefont{L.}~\bibnamefont{Mandel}} \bibnamefont{and}
  \bibinfo{author}{\bibfnamefont{E.}~\bibnamefont{Wolf}},
  \emph{\bibinfo{title}{Optical Coherence and Quantum Optics}}
  (\bibinfo{publisher}{Cambridge University Press}, \bibinfo{year}{1995}).

\bibitem[{\citenamefont{Hastings and Koma}(2006)}]{Hastings2006}
\bibinfo{author}{\bibfnamefont{M.}~\bibnamefont{Hastings}} \bibnamefont{and}
  \bibinfo{author}{\bibfnamefont{T.}~\bibnamefont{Koma}},
  \bibinfo{journal}{Commun. Math. Phys.} \textbf{\bibinfo{volume}{265}},
  \bibinfo{pages}{781} (\bibinfo{year}{2006}).

\bibitem[{\citenamefont{Sachdev}(2000)}]{Sachdev2000}
\bibinfo{author}{\bibfnamefont{S.}~\bibnamefont{Sachdev}},
  \emph{\bibinfo{title}{Quantum Phase Transitions}}
  (\bibinfo{publisher}{Cambridge University Press}, \bibinfo{year}{2000}).

\bibitem[{\citenamefont{K\"uhner and Monien}(1998)}]{Kuehner1998}
\bibinfo{author}{\bibfnamefont{T.~D.} \bibnamefont{K\"uhner}} \bibnamefont{and}
  \bibinfo{author}{\bibfnamefont{H.}~\bibnamefont{Monien}},
  \bibinfo{journal}{Phys. Rev. B} \textbf{\bibinfo{volume}{58}},
  \bibinfo{pages}{R14741} (\bibinfo{year}{1998}).

\bibitem[{\citenamefont{Popov}(1987)}]{Popov1987}
\bibinfo{author}{\bibfnamefont{V.~N.} \bibnamefont{Popov}},
  \emph{\bibinfo{title}{Functional integrals and collective excitations}}
  (\bibinfo{publisher}{Cambridge University Press}, \bibinfo{year}{1987}).

\bibitem[{\citenamefont{Olshanii}(1998)}]{Olshanii1998}
\bibinfo{author}{\bibfnamefont{M.}~\bibnamefont{Olshanii}},
  \bibinfo{journal}{Phys. Rev. Lett.} \textbf{\bibinfo{volume}{81}},
  \bibinfo{pages}{938} (\bibinfo{year}{1998}).

\bibitem[{\citenamefont{Yukalov and Girardeau}(2005)}]{Yukalov2005}
\bibinfo{author}{\bibfnamefont{V.}~\bibnamefont{Yukalov}} \bibnamefont{and}
  \bibinfo{author}{\bibfnamefont{M.}~\bibnamefont{Girardeau}},
  \bibinfo{journal}{Laser Phys. Lett.} \textbf{\bibinfo{volume}{2}},
  \bibinfo{pages}{375} (\bibinfo{year}{2005}).

\bibitem[{\citenamefont{Cazalilla et~al.}(2011)\citenamefont{Cazalilla, Citro,
  Giamarchi, Orignac, and Rigol}}]{Cazalilla2011}
\bibinfo{author}{\bibfnamefont{M.~A.} \bibnamefont{Cazalilla}},
  \bibinfo{author}{\bibfnamefont{R.}~\bibnamefont{Citro}},
  \bibinfo{author}{\bibfnamefont{T.}~\bibnamefont{Giamarchi}},
  \bibinfo{author}{\bibfnamefont{E.}~\bibnamefont{Orignac}}, \bibnamefont{and}
  \bibinfo{author}{\bibfnamefont{M.}~\bibnamefont{Rigol}},
  \bibinfo{journal}{Rev. Mod. Phys.} \textbf{\bibinfo{volume}{83}},
  \bibinfo{pages}{1405} (\bibinfo{year}{2011}).

\bibitem[{\citenamefont{Friedel}(1958)}]{Friedel1958}
\bibinfo{author}{\bibfnamefont{J.}~\bibnamefont{Friedel}},
  \bibinfo{journal}{Nuovo Cimento} \textbf{\bibinfo{volume}{7}},
  \bibinfo{pages}{287} (\bibinfo{year}{1958}).

\bibitem[{\citenamefont{Albuquerque et~al.}(2007)\citenamefont{Albuquerque,
  Alet, Corboz, Dayal, Feiguin, Fuchs, Gamper, Gull, G\"{u}rtler, Honecker
  et~al.}}]{Albuquerque2007}
\bibinfo{author}{\bibfnamefont{A.}~\bibnamefont{Albuquerque}},
  \bibinfo{author}{\bibfnamefont{F.}~\bibnamefont{Alet}},
  \bibinfo{author}{\bibfnamefont{P.}~\bibnamefont{Corboz}},
  \bibinfo{author}{\bibfnamefont{P.}~\bibnamefont{Dayal}},
  \bibinfo{author}{\bibfnamefont{A.}~\bibnamefont{Feiguin}},
  \bibinfo{author}{\bibfnamefont{S.}~\bibnamefont{Fuchs}},
  \bibinfo{author}{\bibfnamefont{L.}~\bibnamefont{Gamper}},
  \bibinfo{author}{\bibfnamefont{E.}~\bibnamefont{Gull}},
  \bibinfo{author}{\bibfnamefont{S.}~\bibnamefont{G\"{u}rtler}},
  \bibinfo{author}{\bibfnamefont{A.}~\bibnamefont{Honecker}},
  \bibnamefont{et~al.}, \bibinfo{journal}{J. Magn. Magn. Mater.}
  \textbf{\bibinfo{volume}{310}}, \bibinfo{pages}{1187 }
  (\bibinfo{year}{2007}).

\bibitem[{\citenamefont{Bauer et~al.}(2011)\citenamefont{Bauer, Carr, Evertz,
  Feiguin, Freire, Fuchs, Gamper, Gukelberger, Gull, G\"{u}rtler
  et~al.}}]{Bauer2011}
\bibinfo{author}{\bibfnamefont{B.}~\bibnamefont{Bauer}},
  \bibinfo{author}{\bibfnamefont{L.~D.} \bibnamefont{Carr}},
  \bibinfo{author}{\bibfnamefont{H.~G.} \bibnamefont{Evertz}},
  \bibinfo{author}{\bibfnamefont{A.}~\bibnamefont{Feiguin}},
  \bibinfo{author}{\bibfnamefont{J.}~\bibnamefont{Freire}},
  \bibinfo{author}{\bibfnamefont{S.}~\bibnamefont{Fuchs}},
  \bibinfo{author}{\bibfnamefont{L.}~\bibnamefont{Gamper}},
  \bibinfo{author}{\bibfnamefont{J.}~\bibnamefont{Gukelberger}},
  \bibinfo{author}{\bibfnamefont{E.}~\bibnamefont{Gull}},
  \bibinfo{author}{\bibfnamefont{S.}~\bibnamefont{G\"{u}rtler}},
  \bibnamefont{et~al.}, \bibinfo{journal}{J. Stat. Mech. Theor. Exp.}
  \textbf{\bibinfo{volume}{2011}}, \bibinfo{pages}{P05001}
  (\bibinfo{year}{2011}).

\bibitem[{\citenamefont{White}(1993)}]{White1993}
\bibinfo{author}{\bibfnamefont{S.~R.} \bibnamefont{White}},
  \bibinfo{journal}{Phys. Rev. B} \textbf{\bibinfo{volume}{48}},
  \bibinfo{pages}{10345} (\bibinfo{year}{1993}).

\bibitem[{\citenamefont{Schollw\"ock}(2005)}]{Schollwock2005}
\bibinfo{author}{\bibfnamefont{U.}~\bibnamefont{Schollw\"ock}},
  \bibinfo{journal}{Rev. Mod. Phys.} \textbf{\bibinfo{volume}{77}},
  \bibinfo{pages}{259} (\bibinfo{year}{2005}).

\bibitem[{\citenamefont{Schollw\"{o}ck}(2011)}]{Schollwock2011}
\bibinfo{author}{\bibfnamefont{U.}~\bibnamefont{Schollw\"{o}ck}},
  \bibinfo{journal}{Ann. Phys.} \textbf{\bibinfo{volume}{326}},
  \bibinfo{pages}{96} (\bibinfo{year}{2011}).

\bibitem[{\citenamefont{K\"uhner et~al.}(2000)\citenamefont{K\"uhner, White,
  and Monien}}]{Kuehner2000}
\bibinfo{author}{\bibfnamefont{T.~D.} \bibnamefont{K\"uhner}},
  \bibinfo{author}{\bibfnamefont{S.~R.} \bibnamefont{White}}, \bibnamefont{and}
  \bibinfo{author}{\bibfnamefont{H.}~\bibnamefont{Monien}},
  \bibinfo{journal}{Phys. Rev. B} \textbf{\bibinfo{volume}{61}},
  \bibinfo{pages}{12474} (\bibinfo{year}{2000}).

\bibitem[{\citenamefont{Ejima et~al.}(2011)\citenamefont{Ejima, Fehske, and
  Gebhard}}]{Ejima2011}
\bibinfo{author}{\bibfnamefont{S.}~\bibnamefont{Ejima}},
  \bibinfo{author}{\bibfnamefont{H.}~\bibnamefont{Fehske}}, \bibnamefont{and}
  \bibinfo{author}{\bibfnamefont{F.}~\bibnamefont{Gebhard}},
  \bibinfo{journal}{Europhys. Lett.} \textbf{\bibinfo{volume}{93}},
  \bibinfo{pages}{30002} (\bibinfo{year}{2011}).

\bibitem[{\citenamefont{Kosterlitz and Thouless}(1973)}]{Kosterlitz1973}
\bibinfo{author}{\bibfnamefont{J.~M.} \bibnamefont{Kosterlitz}}
  \bibnamefont{and} \bibinfo{author}{\bibfnamefont{D.~J.}
  \bibnamefont{Thouless}}, \bibinfo{journal}{J. Phys. C}
  \textbf{\bibinfo{volume}{6}}, \bibinfo{pages}{1181} (\bibinfo{year}{1973}).

\bibitem[{\citenamefont{Pai et~al.}(1996)\citenamefont{Pai, Pandit,
  Krishnamurthy, and Ramasesha}}]{Pai1996}
\bibinfo{author}{\bibfnamefont{R.~V.} \bibnamefont{Pai}},
  \bibinfo{author}{\bibfnamefont{R.}~\bibnamefont{Pandit}},
  \bibinfo{author}{\bibfnamefont{H.~R.} \bibnamefont{Krishnamurthy}},
  \bibnamefont{and}
  \bibinfo{author}{\bibfnamefont{S.}~\bibnamefont{Ramasesha}},
  \bibinfo{journal}{Phys. Rev. Lett.} \textbf{\bibinfo{volume}{76}},
  \bibinfo{pages}{2937} (\bibinfo{year}{1996}).

\bibitem[{\citenamefont{Ejima et~al.}(2012)\citenamefont{Ejima, Fehske,
  Gebhard, zu~M\"{u}nster, Knap, Anrrigoni, and von~der Linden}}]{Ejima2012}
\bibinfo{author}{\bibfnamefont{S.}~\bibnamefont{Ejima}},
  \bibinfo{author}{\bibfnamefont{H.}~\bibnamefont{Fehske}},
  \bibinfo{author}{\bibfnamefont{F.}~\bibnamefont{Gebhard}},
  \bibinfo{author}{\bibfnamefont{K.}~\bibnamefont{zu~M\"{u}nster}},
  \bibinfo{author}{\bibfnamefont{M.}~\bibnamefont{Knap}},
  \bibinfo{author}{\bibfnamefont{E.}~\bibnamefont{Anrrigoni}},
  \bibnamefont{and} \bibinfo{author}{\bibfnamefont{W.}~\bibnamefont{von~der
  Linden}}, \bibinfo{journal}{Phys. Rev. A} \textbf{\bibinfo{volume}{85}},
  \bibinfo{pages}{053644} (\bibinfo{year}{2012}).

\bibitem[{\citenamefont{Kollath et~al.}(2004)\citenamefont{Kollath,
  Schollw\"ock, von Delft, and Zwerger}}]{Kollath2004}
\bibinfo{author}{\bibfnamefont{C.}~\bibnamefont{Kollath}},
  \bibinfo{author}{\bibfnamefont{U.}~\bibnamefont{Schollw\"ock}},
  \bibinfo{author}{\bibfnamefont{J.}~\bibnamefont{von Delft}},
  \bibnamefont{and} \bibinfo{author}{\bibfnamefont{W.}~\bibnamefont{Zwerger}},
  \bibinfo{journal}{Phys. Rev. A} \textbf{\bibinfo{volume}{69}},
  \bibinfo{pages}{031601} (\bibinfo{year}{2004}).

\bibitem[{\citenamefont{Fisher et~al.}(1973)\citenamefont{Fisher, Barber, and
  Jasnow}}]{Fisher1973}
\bibinfo{author}{\bibfnamefont{M.~E.} \bibnamefont{Fisher}},
  \bibinfo{author}{\bibfnamefont{M.~N.} \bibnamefont{Barber}},
  \bibnamefont{and} \bibinfo{author}{\bibfnamefont{D.}~\bibnamefont{Jasnow}},
  \bibinfo{journal}{Phys. Rev. A} \textbf{\bibinfo{volume}{8}},
  \bibinfo{pages}{1111} (\bibinfo{year}{1973}).

\bibitem[{\citenamefont{Krauth}(1991)}]{Krauth1991}
\bibinfo{author}{\bibfnamefont{W.}~\bibnamefont{Krauth}},
  \bibinfo{journal}{Phys. Rev. B} \textbf{\bibinfo{volume}{44}},
  \bibinfo{pages}{9772} (\bibinfo{year}{1991}).

\bibitem[{\citenamefont{Singh and Rokhsar}(1994)}]{Singh1994}
\bibinfo{author}{\bibfnamefont{K.~G.} \bibnamefont{Singh}} \bibnamefont{and}
  \bibinfo{author}{\bibfnamefont{D.~S.} \bibnamefont{Rokhsar}},
  \bibinfo{journal}{Phys. Rev. B} \textbf{\bibinfo{volume}{49}},
  \bibinfo{pages}{9013} (\bibinfo{year}{1994}).

\bibitem[{\citenamefont{Roth and Burnett}(2003{\natexlab{a}})}]{Roth2003}
\bibinfo{author}{\bibfnamefont{R.}~\bibnamefont{Roth}} \bibnamefont{and}
  \bibinfo{author}{\bibfnamefont{K.}~\bibnamefont{Burnett}},
  \bibinfo{journal}{Phys. Rev. A} \textbf{\bibinfo{volume}{68}},
  \bibinfo{pages}{023604} (\bibinfo{year}{2003}{\natexlab{a}}).

\bibitem[{\citenamefont{Roth and Burnett}(2003{\natexlab{b}})}]{Roth2003a}
\bibinfo{author}{\bibfnamefont{R.}~\bibnamefont{Roth}} \bibnamefont{and}
  \bibinfo{author}{\bibfnamefont{K.}~\bibnamefont{Burnett}},
  \bibinfo{journal}{Phys. Rev. A} \textbf{\bibinfo{volume}{67}},
  \bibinfo{pages}{031602} (\bibinfo{year}{2003}{\natexlab{b}}).

\bibitem[{\citenamefont{Penrose and Onsager}(1956)}]{Penrose1956}
\bibinfo{author}{\bibfnamefont{O.}~\bibnamefont{Penrose}} \bibnamefont{and}
  \bibinfo{author}{\bibfnamefont{L.}~\bibnamefont{Onsager}},
  \bibinfo{journal}{Phys. Rev.} \textbf{\bibinfo{volume}{104}},
  \bibinfo{pages}{576} (\bibinfo{year}{1956}).

\bibitem[{\citenamefont{Penrose}(1951)}]{Penrose1951}
\bibinfo{author}{\bibfnamefont{O.}~\bibnamefont{Penrose}},
  \bibinfo{journal}{Philos. Mag.} \textbf{\bibinfo{volume}{42}},
  \bibinfo{pages}{1373} (\bibinfo{year}{1951}).

\bibitem[{\citenamefont{Li and Haldane}(2008)}]{Li2008b}
\bibinfo{author}{\bibfnamefont{H.}~\bibnamefont{Li}} \bibnamefont{and}
  \bibinfo{author}{\bibfnamefont{F.~D.~M.} \bibnamefont{Haldane}},
  \bibinfo{journal}{Phys. Rev. Lett.} \textbf{\bibinfo{volume}{101}},
  \bibinfo{pages}{010504} (\bibinfo{year}{2008}).

\bibitem[{\citenamefont{Spekkens and Sipe}(1999)}]{Spekkens1999}
\bibinfo{author}{\bibfnamefont{R.~W.} \bibnamefont{Spekkens}} \bibnamefont{and}
  \bibinfo{author}{\bibfnamefont{J.~E.} \bibnamefont{Sipe}},
  \bibinfo{journal}{Phys. Rev. A} \textbf{\bibinfo{volume}{59}},
  \bibinfo{pages}{3868} (\bibinfo{year}{1999}).

\bibitem[{\citenamefont{Mueller et~al.}(2006)\citenamefont{Mueller, Ho, Ueda,
  and Baym}}]{Mueller2006}
\bibinfo{author}{\bibfnamefont{E.~J.} \bibnamefont{Mueller}},
  \bibinfo{author}{\bibfnamefont{T.-L.} \bibnamefont{Ho}},
  \bibinfo{author}{\bibfnamefont{M.}~\bibnamefont{Ueda}}, \bibnamefont{and}
  \bibinfo{author}{\bibfnamefont{G.}~\bibnamefont{Baym}},
  \bibinfo{journal}{Phys. Rev. A} \textbf{\bibinfo{volume}{74}},
  \bibinfo{pages}{033612} (\bibinfo{year}{2006}).

\bibitem[{\citenamefont{Nielsen and Chuang}(2000)}]{Nielsen2000}
\bibinfo{author}{\bibfnamefont{M.~A.} \bibnamefont{Nielsen}} \bibnamefont{and}
  \bibinfo{author}{\bibfnamefont{I.~L.} \bibnamefont{Chuang}},
  \emph{\bibinfo{title}{Quantum Computation and Quantum Information}}
  (\bibinfo{publisher}{Cambridge University Press}, \bibinfo{year}{2000}).

\bibitem[{\citenamefont{Peschel and Eisler}(2009)}]{Peschel2009}
\bibinfo{author}{\bibfnamefont{I.}~\bibnamefont{Peschel}} \bibnamefont{and}
  \bibinfo{author}{\bibfnamefont{V.}~\bibnamefont{Eisler}},
  \bibinfo{journal}{J. Phys. A: Math. Theor.} \textbf{\bibinfo{volume}{42}},
  \bibinfo{pages}{504003} (\bibinfo{year}{2009}).

\bibitem[{\citenamefont{Eisert et~al.}(2010)\citenamefont{Eisert, Cramer, and
  Plenio}}]{Eisert2010}
\bibinfo{author}{\bibfnamefont{J.}~\bibnamefont{Eisert}},
  \bibinfo{author}{\bibfnamefont{M.}~\bibnamefont{Cramer}}, \bibnamefont{and}
  \bibinfo{author}{\bibfnamefont{M.~B.} \bibnamefont{Plenio}},
  \bibinfo{journal}{Rev. Mod. Phys.} \textbf{\bibinfo{volume}{82}},
  \bibinfo{pages}{277} (\bibinfo{year}{2010}).

\bibitem[{\citenamefont{Vidal}(2003)}]{Vidal2003}
\bibinfo{author}{\bibfnamefont{G.}~\bibnamefont{Vidal}},
  \bibinfo{journal}{Phys. Rev. Lett.} \textbf{\bibinfo{volume}{91}},
  \bibinfo{pages}{147902} (\bibinfo{year}{2003}).

\bibitem[{\citenamefont{Wolf}(2006)}]{Wolf2006}
\bibinfo{author}{\bibfnamefont{M.~M.} \bibnamefont{Wolf}},
  \bibinfo{journal}{Phys. Rev. Lett.} \textbf{\bibinfo{volume}{96}},
  \bibinfo{pages}{010404} (\bibinfo{year}{2006}).

\bibitem[{\citenamefont{Plenio et~al.}(2005)\citenamefont{Plenio, Eisert,
  Dreissig, and Cramer}}]{Plenio2005}
\bibinfo{author}{\bibfnamefont{M.~B.} \bibnamefont{Plenio}},
  \bibinfo{author}{\bibfnamefont{J.}~\bibnamefont{Eisert}},
  \bibinfo{author}{\bibfnamefont{J.}~\bibnamefont{Dreissig}}, \bibnamefont{and}
  \bibinfo{author}{\bibfnamefont{M.}~\bibnamefont{Cramer}},
  \bibinfo{journal}{Phys. Rev. Lett.} \textbf{\bibinfo{volume}{94}},
  \bibinfo{pages}{060503} (\bibinfo{year}{2005}).

\bibitem[{\citenamefont{Van~den Nest}(2013)}]{VanDenNest2013}
\bibinfo{author}{\bibfnamefont{M.}~\bibnamefont{Van~den Nest}},
  \bibinfo{journal}{Phys. Rev. Lett.} \textbf{\bibinfo{volume}{110}},
  \bibinfo{pages}{060504} (\bibinfo{year}{2013}).

\bibitem[{\citenamefont{Verstraete
  et~al.}(2004{\natexlab{a}})\citenamefont{Verstraete, Popp, and
  Cirac}}]{Verstraete2004c}
\bibinfo{author}{\bibfnamefont{F.}~\bibnamefont{Verstraete}},
  \bibinfo{author}{\bibfnamefont{M.}~\bibnamefont{Popp}}, \bibnamefont{and}
  \bibinfo{author}{\bibfnamefont{J.~I.} \bibnamefont{Cirac}},
  \bibinfo{journal}{Phys. Rev. Lett.} \textbf{\bibinfo{volume}{92}},
  \bibinfo{pages}{027901} (\bibinfo{year}{2004}{\natexlab{a}}).

\bibitem[{\citenamefont{Verstraete
  et~al.}(2004{\natexlab{b}})\citenamefont{Verstraete, Mart\'{\i}n-Delgado, and
  Cirac}}]{Verstraete2004b}
\bibinfo{author}{\bibfnamefont{F.}~\bibnamefont{Verstraete}},
  \bibinfo{author}{\bibfnamefont{M.~A.} \bibnamefont{Mart\'{\i}n-Delgado}},
  \bibnamefont{and} \bibinfo{author}{\bibfnamefont{J.~I.} \bibnamefont{Cirac}},
  \bibinfo{journal}{Phys. Rev. Lett.} \textbf{\bibinfo{volume}{92}},
  \bibinfo{pages}{087201} (\bibinfo{year}{2004}{\natexlab{b}}).

\bibitem[{\citenamefont{Popp et~al.}(2005)\citenamefont{Popp, Verstraete,
  Mart\'\i{}n-Delgado, and Cirac}}]{Popp2005}
\bibinfo{author}{\bibfnamefont{M.}~\bibnamefont{Popp}},
  \bibinfo{author}{\bibfnamefont{F.}~\bibnamefont{Verstraete}},
  \bibinfo{author}{\bibfnamefont{M.~A.} \bibnamefont{Mart\'\i{}n-Delgado}},
  \bibnamefont{and} \bibinfo{author}{\bibfnamefont{J.~I.} \bibnamefont{Cirac}},
  \bibinfo{journal}{Phys. Rev. A} \textbf{\bibinfo{volume}{71}},
  \bibinfo{pages}{042306} (\bibinfo{year}{2005}).

\bibitem[{\citenamefont{Amico et~al.}(2008)\citenamefont{Amico, Fazio,
  Osterloh, and Vedral}}]{Amico2008}
\bibinfo{author}{\bibfnamefont{L.}~\bibnamefont{Amico}},
  \bibinfo{author}{\bibfnamefont{R.}~\bibnamefont{Fazio}},
  \bibinfo{author}{\bibfnamefont{A.}~\bibnamefont{Osterloh}}, \bibnamefont{and}
  \bibinfo{author}{\bibfnamefont{V.}~\bibnamefont{Vedral}},
  \bibinfo{journal}{Rev. Mod. Phys.} \textbf{\bibinfo{volume}{80}},
  \bibinfo{pages}{517} (\bibinfo{year}{2008}).

\bibitem[{\citenamefont{Carmichael et~al.}(1996)\citenamefont{Carmichael,
  Kochan, and Sanders}}]{Carmichael1996}
\bibinfo{author}{\bibfnamefont{H.~J.} \bibnamefont{Carmichael}},
  \bibinfo{author}{\bibfnamefont{P.}~\bibnamefont{Kochan}}, \bibnamefont{and}
  \bibinfo{author}{\bibfnamefont{B.~C.} \bibnamefont{Sanders}},
  \bibinfo{journal}{Phys. Rev. Lett.} \textbf{\bibinfo{volume}{77}},
  \bibinfo{pages}{631} (\bibinfo{year}{1996}).

\bibitem[{\citenamefont{Zhu et~al.}(2013)\citenamefont{Zhu, Schmidt, and
  Koch}}]{Zhu2013}
\bibinfo{author}{\bibfnamefont{G.}~\bibnamefont{Zhu}},
  \bibinfo{author}{\bibfnamefont{S.}~\bibnamefont{Schmidt}}, \bibnamefont{and}
  \bibinfo{author}{\bibfnamefont{J.}~\bibnamefont{Koch}},
  \textbf{\bibinfo{volume}{arXiv:1307.2505}} (\bibinfo{year}{2013}).

\bibitem[{\citenamefont{Vidal}(2008)}]{Vidal2008}
\bibinfo{author}{\bibfnamefont{G.}~\bibnamefont{Vidal}},
  \bibinfo{journal}{Phys. Rev. Lett.} \textbf{\bibinfo{volume}{101}},
  \bibinfo{pages}{110501} (\bibinfo{year}{2008}).

\bibitem[{\citenamefont{McCulloch}(2008)}]{McCulloch2008}
\bibinfo{author}{\bibfnamefont{I.~P.} \bibnamefont{McCulloch}},
  \textbf{\bibinfo{volume}{arXiv:0804.2509}} (\bibinfo{year}{2008}).

\bibitem[{\citenamefont{Crosswhite et~al.}(2008)\citenamefont{Crosswhite,
  Doherty, and Vidal}}]{Crosswhite2008}
\bibinfo{author}{\bibfnamefont{G.~M.} \bibnamefont{Crosswhite}},
  \bibinfo{author}{\bibfnamefont{A.~C.} \bibnamefont{Doherty}},
  \bibnamefont{and} \bibinfo{author}{\bibfnamefont{G.}~\bibnamefont{Vidal}},
  \bibinfo{journal}{Phys. Rev. B} \textbf{\bibinfo{volume}{78}},
  \bibinfo{pages}{035116} (\bibinfo{year}{2008}).

\bibitem[{\citenamefont{Briegel et~al.}(2009)\citenamefont{Briegel, Browne,
  D{\"u}r, Raussendorf, and Van~den Nest}}]{Briegel2009}
\bibinfo{author}{\bibfnamefont{H.~J.} \bibnamefont{Briegel}},
  \bibinfo{author}{\bibfnamefont{D.~E.} \bibnamefont{Browne}},
  \bibinfo{author}{\bibfnamefont{W.}~\bibnamefont{D{\"u}r}},
  \bibinfo{author}{\bibfnamefont{R.}~\bibnamefont{Raussendorf}},
  \bibnamefont{and} \bibinfo{author}{\bibfnamefont{M.}~\bibnamefont{Van~den
  Nest}}, \bibinfo{journal}{Nat. Phys.} \textbf{\bibinfo{volume}{5}},
  \bibinfo{pages}{19} (\bibinfo{year}{2009}).

\end{thebibliography}

\end{document}